\documentclass{article}

\usepackage{arxiv}

\usepackage[utf8]{inputenc} 
\usepackage[T1]{fontenc}    
\usepackage{hyperref}       
\usepackage{url}            
\usepackage{booktabs}       
\usepackage{amsfonts}       
\usepackage{nicefrac}       
\usepackage{microtype}      
\usepackage{lipsum}
\usepackage{cancel}
\usepackage{bm}
\usepackage{nicefrac}
\usepackage{hyperref}
\usepackage{graphicx}

\usepackage{amsmath}
\usepackage[colorinlistoftodos]{todonotes}
\DeclareMathOperator*{\argminC}{\arg\min}   
\usepackage{subcaption}
\newcommand{\rvect}[1]{\begin{bmatrix} #1 \end{bmatrix}}
\usepackage{xcolor}
\newcommand{\code}[1]{{\texttt{#1}}}

 \usepackage{soul} 
 
 \definecolor{red}{rgb}{1,0,0}
 \definecolor{gre}{rgb}{0,1,0}
 \definecolor{blu}{rgb}{0,0,1}

\title{Optimized finite-build stellarator coils using automatic differentiation}

\author{
  Nick McGreivy \\
  Department of Astrophysical Sciences\\
  Princeton University\\
  Princeton, NJ \\
  \texttt{mcgreivy@princeton.edu} \\
   \And
 Stuart R. Hudson \\
  Princeton Plasma Physics Laboratory\\
  Princeton University \\
  PO Box 451, Princeton NJ 08543 \\
  \texttt{shudson@pppl.gov} \\
     \And
  Caoxiang Zhu \\
  Princeton Plasma Physics Laboratory\\
  Princeton University \\
  PO Box 451, Princeton NJ 08543\\
  \texttt{czhu@pppl.gov} \\

}

\renewcommand{\headeright}{}
\renewcommand{\undertitle}{}
\renewcommand{\shorttitle}{}

\hypersetup{
pdftitle={Finite build stellarator coil design},
pdfsubject={physics.plasm-ph},
pdfauthor={Nick McGreivy, Stuart Hudson. Caoxiang Zhu},
pdfkeywords={Stellarator Optimization, Automatic Differentiation, Stellarator, Coil Design},
}
 \newcommand{\edit}[1]{#1}

\begin{document}
\maketitle

\begin{abstract}

A new stellarator coil design code is introduced that optimizes the position and winding pack \edit{orientation} of finite-build coils. The new code, called FOCUSADD, performs gradient-based optimization in a high-dimensional, non-convex space. The derivatives with respect to parameters of finite-build coils are easily and efficiently computed using automatic differentiation. 
FOCUSADD parametrizes coil positions in free space using a Fourier series and uses a multi-filament approximation to the coil winding pack. The \edit{orientation} of the winding pack is parametrized with a Fourier series and can be optimized as well. Optimized finite-build coils for a W7-X\edit{-like} stellarator are found, and compared with filamentary coil results. The final positions of optimized finite-build W7-X\edit{-like} coils are shifted, on average, by approximately 2.5mm relative to optimized filamentary coils. These results suggest that finite-build effects should be accounted for in the optimization of stellarators with low coil tolerances.
\end{abstract}



\section{Introduction}

The stellarator is a toroidal magnetic fusion concept which confines plasma using a rotational transform of the vacuum magnetic field \cite{Helander_2014}. 
The rotational transform of the vacuum magnetic field in a stellarator device is created by non-axisymmetric current-carrying coils. The non-axisymmetry of the current-carrying coils and magnetic field allow for a large number of degrees of freedom in the design of a stellarator device. Stellarator design therefore can be formulated as an optimization problem over these degrees of freedom \cite{boozer_C02}. The objective of this optimization problem is to simultaneously maximize the plasma performance and minimize the engineering and construction costs of the device. 

Well-designed coils are prerequisites to achieving the performance and cost goals of a stellarator device. 
Usually, stellarator coils are designed to reproduce a given target magnetic field. 
Because inverting the Biot-Savart law is an ill-posed problem, there is no coil set with a finite number of coils that can exactly reproduce an arbitrary magnetic field \edit{throughout a volume}.
Therefore, the goal of stellarator coil design is to find a set of coils which produces the target magnetic field well enough to accomplish the performance goals of the experiment and which can be built and assembled at the cheapest possible cost. 

In practice, achieving the goals of stellarator coil design means successfully optimizing a well-crafted objective function which includes a number of complex physics and engineering objectives. \edit{One term in this objective function usually encourages minimizing the surface integral of the normal magnetic field squared on the outer plasma surface. Other terms could be added; f}or example, a\edit{n} objective function \edit{could be crafted which gives} resonant error fields greater weight in the objective function relative to less damaging non-resonant error fields. \edit{Explicitly targeting resonant error fields was implemented in the island healing techniques developed for reducing chaos in the NCSX stellarator \cite{PhysRevLett.89.275003}.
More recently, a method for identifying the important error fields was presented by Zhu {\em et al.} \cite{Zhu_2019}.} In addition, the coil-coil spacing should be as large as possible to allow for increased access to the plasma for maintenance, diagnostics, and beams. In a power plant, the coils should be at least one meter from the plasma to allow for the tritium breeding blanket and neutron shielding. Each of these objectives could be converted to a potentially complex scalar objective function or possibly a constraint on the optimization which allows for a set of coils to be found which best satisfy the desired engineering constraints.  

Successful optimization of a scalar objective function in non-convex, high-dimensional spaces often relies on the use of derivative information to inform the optimization process. Many stellarator coil design codes have performed gradient-based optimization by computing finite-difference derivatives \cite{strickler_2002}, while recent work has allowed for efficient derivative computations by computing analytic derivatives \cite{Paul_2018,Zhu_2018,paul_abel_landreman_dorland_2019,antonsen_paul_landreman_2019}. In this work, we instead use automatic differentiation to efficiently compute the required derivatives. 

\begin{figure}
    \centering
    \includegraphics[scale=0.5]{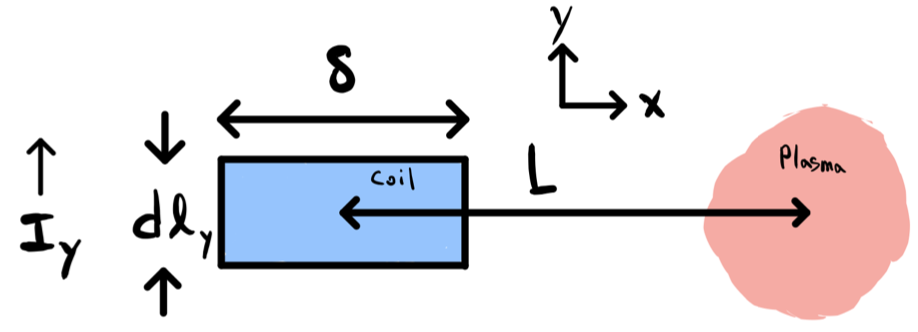}
    \caption{A sketch of an infinitesimal finite-build coil segment carrying current in the $y$-direction. The coil is assumed to have zero thickness in the $z$-direction, and thickness $\delta$ in the $x$-direction. The center of the coil is a distance $L$ from the plasma. The lowest-order correction of the magnetic field $B_z$ due to the finite-build of the coil is second-order in $\delta/L$. }
    \label{fig:biot-savart}
\end{figure}

Existing coil design codes have ultimately optimized the positions of filamentary (zero thickness) coils. However, any real coil will be made up of a winding pack that carries current over a non-zero volume. While such a coil can be approximated as a single filament in space, this approximation is only valid in the limit that the coil thickness is much less than the distance from the coil to the plasma. The correction to the magnetic field due to coil finite build is second-order in the coil thickness $\delta$ divided by the coil-plasma distance $L$, which can be shown using a Taylor expansion of the magnetic field due to an infinitesimal finite-build coil segment as sketched in figure \ref{fig:biot-savart}. Assuming the coil has zero thickness in the $z$-direction, the magnetic field at the plasma is given by
\begin{equation}\label{eq:one}
        \mathrm{d}B_z = -\frac{\mu_0 I_y \mathrm{d} \ell_y}{4 \pi \delta }\int_{-\delta/2}^{\delta/2} \frac{\mathrm{d}x}{(L+x)^2}\edit{.}
\end{equation}
Taylor expanding the denominator to lowest order gives a second-order correction in the ratio $\nicefrac{\delta}{L}$:
\begin{equation}\label{eq:expansion}
    \mathrm{d}B_z \approx -\frac{\mu_0 I_y \mathrm{d}\ell_y}{4\pi \delta L^2}\int_{-\delta/2}^{\delta/2} \Big(1 - \cancelto{0}{\frac{2x}{L}} + \frac{3x^2}{L^2}\Big)\mathrm{d}x \approx \mathrm{d}B_{\text{filament}}\Big(1 + \frac{\delta^2}{4L^2}\Big)\edit{.}
\end{equation}

\edit{Previous experiments have constructed the geometry of finite-build coils in different ways. One approach, used by the Wendelstein 7-X (W7-X) experiment, is to define an orthonormal frame around a central filament consisting of the tangent to the filament, the normal to a winding surface, and the binormal perpendicular to both, and then defining the edges of rectangular cross-section coils by extending the edges outwards from the central filament in the normal and binormal directions \cite{physics_engineering_design_w7x}.}

The main point of this paper is to introduce a new stellarator coil design code which directly optimizes finite-build (non-zero thickness) coils, thereby accounting for the finite-build correction to the produced magnetic field. This code is called FOCUSADD (\textbf{F}lexible \textbf{O}ptimized \textbf{Cu}rves in \textbf{S}pace using \textbf{A}utomatic \textbf{D}ifferentiation and Finite Buil\textbf{d}). 
A secondary purpose of this paper is to introduce automatic differentiation to the stellarator community. The results of this paper (Section 5) focus on comparing optimized finite-build coils to optimized filamentary coils. 

Recent work using the OMIC code \cite{l_singh_preprint} has, for the first time, optimized coils which have a finite build. OMIC starts with a set of filamentary coils, defines a winding pack around each coil, then optimizes the \edit{orientation} of the coil winding packs using gradient-based optimization and gradients computed using finite-difference. While FOCUS \cite{FOCUS_Zhu} optimizes the position of filamentary coils and OMIC optimizes the \edit{orientation} of finite-build coils, neither code can optimize both position and \edit{winding pack orientation} at once. Both quantities can be optimized at once by FOCUSADD. While we demonstrate this capability (Section 5), optimizing the \edit{winding pack orientation} of finite-build coils is not a primary focus of this paper. 

OMIC uses finite-difference derivatives to compute the gradient of a scalar objective function. One limitation of this approach is that the computational cost of computing the gradient increases linearly with the number of optimization parameters. Analytic derivatives could, in principle, be used to efficiently compute the required gradient, at a computational cost independent of the number of optimization parameters. However, finding and computing analytic derivatives for finite-build stellarator coils is challenging due to the significantly increased complexity of the objective function. Both of these challenges -- computational efficiency and computing analytic derivatives -- are solved in this paper by using reverse mode automatic differentiation (Section 2) to develop a new finite-build coil design code (Section 3).

Automatic differentiation (AD, Section 2), also known as algorithmic differentiation or computational differentiation, is a method of efficiently computing the exact derivatives to any order of a differentiable function specified by a computer program. AD is a broadly applicable numerical technique, but is particularly useful for gradient-based optimization in high-dimensional spaces. This is because with reverse mode AD, the gradient of a scalar function of $N$ inputs can be computed at a small multiple of the cost of computing the original function, independent of $N$. AD has been used successfully in a number of areas, including machine learning \cite{ADinML,NIPS2015_5954}, engineering design optimization \cite{racecar}, beam physics \cite{beam,Berz1998VIo}, optimal control \cite{optimal_control}, atmospheric science \cite{mitgcm}, biomagnetic inverse problems \cite{biomagnetic}, a plasma edge code \cite{AD_in_plasma_edge}, and computational finance \cite{greeks}.

Automatic differentiation is particularly well-suited to the problem of stellarator coil design. This is because in order to satisfy the many physics and engineering objectives of stellarator coil design, we need to optimize a high-dimensional (\edit{200}-5000 optimization parameters) objective function (Section 4) whose analytic derivatives may be extremely difficult to write down and program analytically. AD allows us to efficiently perform gradient-based optimization of such an objective function; efficiently computing the required derivatives would be a significant challenge if AD were not used. We would also like to be able to efficiently optimize many different objective functions, without taking the time and effort to derive and program analytic derivatives for each objective function that we might consider. Automatic differentiation allows us to neither derive nor program these analytic derivatives; we only need to compute the value of an objective function and its derivatives are computed automatically and efficiently.

\section{Automatic Differentiation}

AD is a family of techniques for computing the exact numerical derivatives of a differentiable function represented by a computer program. AD has been extensively studied, and a number of textbooks and review papers exist on the subject \cite{ADinML,evaluatingDerivatives,adimplementation,uwebook,advancesinad}.

While AD can be used to compute numerical derivatives to any order, in many applications only first-order derivatives are computed. For a function $\bm{y} = f(\bm{x})$ from $\bm{x} \in \mathbb{R}^n$ to $\bm{y} \in \mathbb{R}^m$, first-order AD can be used to compute the Jacobian $\bm{J} = \frac{\partial \bm{y}}{\partial \bm{x}}$ at a particular value of $\bm{x}$. However, in practice AD tools usually compute the product of the Jacobian with a vector using either forward mode AD or reverse mode AD. 
In forward mode, AD tools compute the product $\dot{\bm{y}} = \bm{J} \dot{\bm{x}} \in \mathbb{R}^m$ of the Jacobian $\bm J \in \mathbb{R}^{m\times n}$ with a vector $\dot{\bm{x}} \in \mathbb{R}^n$. 
In reverse mode, AD tools compute the product $\overline{\bm{x}} = \overline{\bm{y}}^T\bm{J} \in \mathbb{R}^n$ of a vector $\overline{\bm{y}}\in \mathbb{R}^m$ with the Jacobian $\bm J \in \mathbb{R}^{m\times n}$. 
These are called the Jacobian-vector product and vector-Jacobian product, respectively.

AD works on the principle that mathematical functions can be written as compositions of primitive operations; these primitive operations form the building blocks of functions computed by AD tools.
A primitive operation is any function whose derivative of the output of that operation with respect to the input of that operation is known to the AD tool. The derivative of the output with respect to the input of a primitive operation is called the elementary partial derivative or elementary Jacobian matrix; these derivatives are usually computed analytically and must be pre-programmed for each primitive operation in the AD library. The primitive operations can be standard mathematical functions such as \code{divide}, \code{sine}, \code{fft}, and \code{ode\_int}, or they can be user-defined custom operations. An AD tool composes primitive operations together to build a function $f$, then computes the Jacobian of that function by multiplying the elementary Jacobian matrices together as specified by the chain rule. The elementary partial derivatives can be multiplied in any order, and in general choosing the most computationally efficient way to multiply these matrices is an NP-complete problem known as the Jacobian accumulation problem \cite{npcomplete}.

Two ways of multiplying the elementary partial derivatives are forward mode AD and reverse mode AD. Forward mode AD computes the partial derivatives at the same time as the function is being computed, in effect multiplying elementary Jacobian matrices forwards from the beginning of the function to the end. In practice, the result of the computation is the Jacobian-vector product. Reverse mode AD computes the function forwards and then computes the partial derivatives backwards, in effect multiplying elementary Jacobian matrices from the end of the function to the beginning. In practice, the result of the computation is the vector-Jacobian product. Forward mode is sometimes called ``tangent linear mode'' while reverse mode is sometimes called ``adjoint mode''. The word ``adjoint'' is simply derived from the fact that $\bm{\overline{y}}^T \bm{J} = \bm{J}^\dag\bm{\overline{y}}$ where $\bm{J}^\dag$ is the hermitian transpose, or adjoint, of the Jacobian matrix.

For a function $f : \mathbb{R}^n \rightarrow \mathbb{R}^m$ which takes time $\mathcal{O}(1)$ to compute, computing the full Jacobian with forward mode takes time $\mathcal{O}(n)$ and almost no additional memory cost. This can be done by computing $n$ Jacobian-vector products where the vectors are the columns of a $n \times n$ identity matrix. Computing the full Jacobian with reverse mode takes time $\mathcal{O}(m)$, and memory cost proportional to the number of intermediate variables in the computation. This can be done by computing $m$ vector-Jacobian products where the vectors $\bm{\overline{y}}$ are the rows of an $m\times m$ identity matrix. For large computations, the memory cost of reverse mode AD can be extremely large; checkpointing strategies \cite{checkpointing,checkpointingwang} can be used to reduce the memory cost of reverse mode AD at the cost of increased runtime. An important feature of reverse mode AD is that for a scalar function $f: \mathbb{R}^n \rightarrow \mathbb{R}$, the runtime cost of computing the $n$-dimensional gradient is a small multiple of the cost of computing the function itself, independent of $n$. While computing the full Hessian matrix of a scalar function is $\mathcal{O}(n)$ the cost of the original function, Hessian-vector products can be computed in time $\mathcal{O}(1)$ \cite{jax2018github}. These are useful for second-order Hessian-free optimization methods \cite{NumericalOptimizationBook}. The full Hessian matrix is useful for sensitivity analysis and understanding coil tolerances, an important area of research in stellarator coil design \cite{zhu_hessian, Zhu_hessian_2019}.

AD is one method of computing derivatives, other methods include numerical differentiation (finite-difference) and hand-programmed analytic derivatives. Analytic differentiation often results in the fastest derivative computations and, like AD, gives exact derivative values. However, analytic derivatives are both error-prone and time-consuming to calculate and program, unlike AD. Numerical differentiation is simple to program, but results in inexact derivative values due to floating-point precision errors and has a computational cost proportional to the number of inputs $n$. When computing the gradient of a scalar function, reverse mode AD is comparable in speed to analytic differentiation and has a computational cost independent of the number of inputs. Automatic differentiation has significant advantages over other methods of computing derivatives, especially for optimization. The main downsides of using AD are that programs need to be written using an AD software tool, and that for very large computations the memory cost of reverse mode AD can be unwieldy.

For a review of AD software tools and their implementation, see \cite{adimplementation} and \cite{reviewofADimplementation}, and the AD community website\footnote{\edit{www.autodiff.org}}. Originally, FOCUSADD was adapted from FOCUS and written in FORTRAN with the source transformation tool OpenAD/F \cite{OpenAD/F}. FOCUSADD was then rewritten in JIT-compiled Python using \href{https://github.com/google/jax}{JAX} \cite{jax2018github}, a research project under development by Google. JAX is a library for composable transformations of Python+NumPy programs, including automatic differentiation, vectorization, JIT-compilation to GPU/TPU, and SPMD parallelization.

\subsection{Forward and Reverse Mode: An Example}


Let us further examine how automatic differentiation works through the use of an example. We will compute the derivatives of the example function $f$ given by
\begin{equation} \label{eq:examplefunc}
    y = f(x_1,x_2) = \sin(e^{x_1} x_2) + x_1^2/x_2\edit{.}
\end{equation}
Since $f$ is a scalar function of two variables, then forward mode AD should be able to compute the full gradient in two forward mode computations, while reverse mode AD should be able to compute the full gradient in one reverse mode computation.  

\begin{figure}
    \centering
    \includegraphics[scale=0.18]{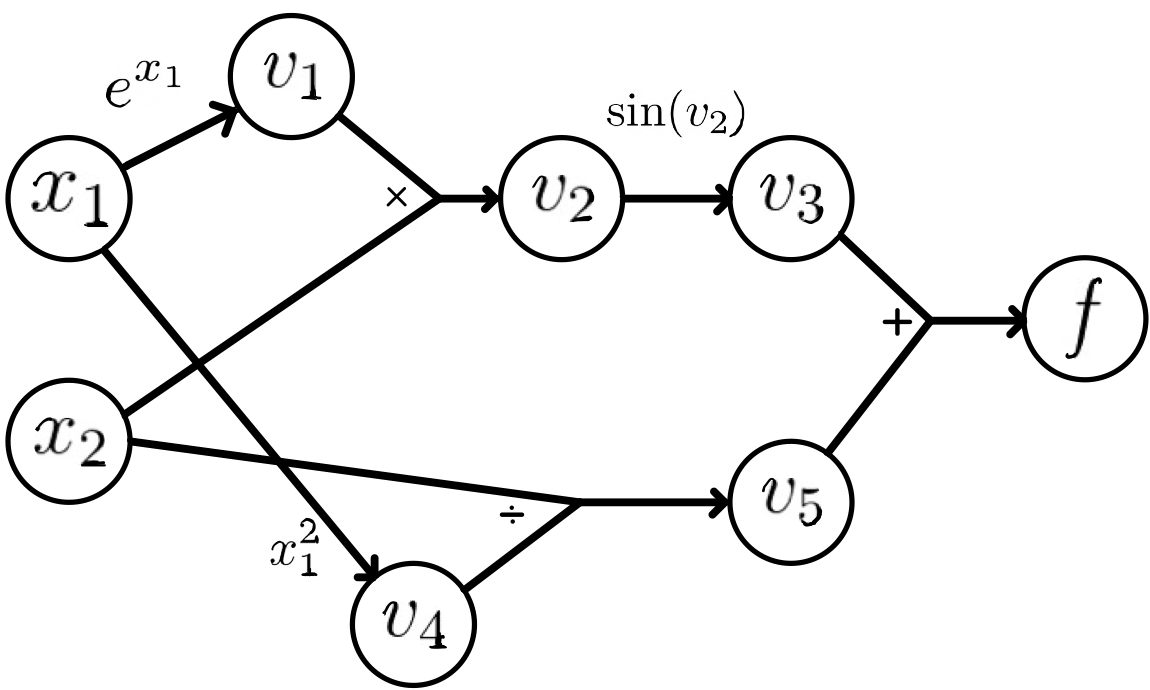}
    \caption{A computational graph for the function defined in equation \ref{eq:examplefunc}. A computational graph is an abstraction commonly used by the AD community to understand the computation performed by a computer program representing a mathematical function. The variables on the left, $x_1$ and $x_2$, are the inputs to the function $f(x_1,x_2)$. The intermediate variables $v_1$, $v_2$, $v_3$, $v_4$, and $v_5$ are produced during the computation of $f$. The function $f$ is computed from left to right across the graph. In forward mode, the derivatives are also computed from left to right across the graph, at the same time as the values are being computed. In reverse mode, the derivatives are computed from right to left after the function is computed from left to right.} 
    \label{fig:compgraphAD}
\end{figure}

A computational graph is an abstraction commonly used by the AD community for understanding the computations performed by a differentiable computer program. A computational graph for $f$ is shown in figure \ref{fig:compgraphAD}. Each vertex in the graph represents a variable used in the computation. For example, the vertex $v_3$ in figure \ref{fig:compgraphAD} represents the intermediate variable which is computed as the result of the \code{sine} function applied to the value in $v_2$. Each edge or combination of edges in the graph represents an elementary operation performed by the function. For example, the edge between $v_2$ and $v_3$ represents the elementary operation which is the \code{sine} function. The edge between $v_3$, $v_5$, and $y$ represents the elementary operation of adding two variables.

\subsubsection{Forward Mode}

We now perform forward mode AD on $f$ to compute the Jacobian-vector product $\bm{J\dot{x}}$ at $x_1=2$ and $x_2=4$. Since $f: \mathbb{R}^2 \rightarrow \mathbb{R}$, then $\bm{J} \in \mathbb{R}^{1\times2}$. Here we will set $\bm{\dot{x}} = \rvect{1,0}^T$ which will give us the partial derivative $\frac{\partial f}{\partial x_1}$ at $x_1=2$ and $x_2=4$:
\begin{equation}
    \dot{y}
     = 
    \begin{bmatrix}
    \frac{\partial f}{\partial x_1} & \frac{\partial f}{\partial x_2} 
    \end{bmatrix}
    \begin{bmatrix} 1 \\
    0
    \end{bmatrix} = \frac{\partial f}{\partial x_1}\edit{.}
\end{equation}

Forward mode AD associates a derivative value $\dot{v}_i$ for each intermediate variable $v_i$. $\dot{v}_i$ is called the tangent variable of $v_i$. $\dot{v}_i$ is computed at the same time as $v_i$, which is equivalent to multiplying the partial derivatives of $f$ from beginning to end. Because $\bm{\dot{x}} = \rvect{1,0}^T$, then the derivative value $\dot{v}_i$ is $\frac{\partial v_i}{\partial x_1}$. If $\bm{\dot{x}}$ were $\rvect{0,1}^T$, then $\dot{v}_i$ would be $\frac{\partial v_i}{\partial x_2}$. If $\bm{\dot{x}}$ were $\rvect{\alpha,\beta}^T$, then $\dot{v}_i$ would be $\alpha \frac{\partial v_i}{\partial x_1} + \beta  \frac{\partial v_i}{\partial x_2}$.

\begin{table}
    \centering
    \begin{tabular}{|c|c|c|c|c|}
        \hline Step &
        Variable & Value & Tangent  &  Value \\ \hline
        1 & $x_1$ & 2.0 & $\dot{x}_1 = \frac{\partial x_1}{\partial x_1}$ & 1.0 \\ \hline
        2 & $x_2$ & 4.0 & $\dot{x}_2 =\frac{\partial x_2}{\partial x_1}$ & 0.0 \\ \hline
        3 & $v_1$ & 7.389 & $\dot{v}_1 =\dot{x}_1 \frac{\partial v_1}{\partial x_1}$ &  7.389  \\ \hline
        4 & $v_2$ & \edit{29.556} & $\dot{v}_2 =\dot{v}_1 \frac{\partial v_2}{\partial v_1} + \dot{x}_2 \frac{\partial v_2}{\partial x_2}$ & \edit{29.556} \\ \hline
        5 & $v_3$ & -0.959 & $\dot{v}_3 =\dot{v}_2 \frac{\partial v_3}{\partial v_2}$ & -8.42 \\ \hline
        6 & $v_4$ & 4.0 & $\dot{v}_4 =\dot{x}_1\frac{\partial v_4}{\partial x_1}$ & 4.0 \\ \hline
        7 & $v_5$ & 1.0 & $\dot{v}_5 =\dot{x}_2 \frac{\partial v_5}{\partial x_2}+ \dot{v}_4 \frac{\partial v_5}{\partial v_4}$ & 1.0 \\ \hline
        8 & $y$ & 0.0414 & $\dot{y} = \dot{v}_3\frac{\partial y}{\partial v_3} + \dot{v}_5 \frac{\partial y}{\partial v_5}$ & -7.421 \\ \hline
    \end{tabular}
    \caption{The steps of the forward mode computation which compute the function from equation \ref{eq:examplefunc} and its derivative $\frac{\partial f}{\partial x_1}$. In steps 1 and 2, the input variables $x_1$ and $x_2$ and their tangents are set. In steps 3-7, the intermediate variables in figure \ref{fig:compgraphAD} are computed and their tangents are computed using the update rule in equation \ref{eq:forwardupdate}. In step 8, the output variable $y$ is set, along with its tangent. }
    \label{tab:forwardmode}
\end{table}

We can use the computational graph in figure \ref{fig:compgraphAD} along with table \ref{tab:forwardmode} to understand how the derivative $\frac{\partial f}{\partial x_1}$ is computed with forward mode AD. In steps 1 and 2 of table \ref{tab:forwardmode}, $x_1$ and $x_2$ are set to $2$ and $4$, while their tangents $\dot{x}_1$ and $\dot{x}_2$ are set to 1 and 0 to match $\bm{\dot{x}} = \rvect{1,0}^T$. In steps 3-8, the computational graph in figure \ref{fig:compgraphAD} is traversed in topological order. For each vertex, $v_i$ is computed along with its tangent $\dot{v}_i$. $\dot{v}_i$ is computed using the update rule
\begin{equation} \label{eq:forwardupdate}
    \dot{v}_i = \sum_{\substack{j \in \text{parents} \\ \text{of } i}} \dot{v}_j \frac{\partial v_i}{\partial v_j}\edit{.}
\end{equation}
\edit{This update rule is equivalent to the chain rule.}

In the last row of table \ref{tab:forwardmode} we have the result of the computation $y = 0.0414$, as well as the tangent $\dot{y} \equiv \frac{\partial f}{\partial x_1} = - 7.421$. Notice that this gave us one column of the Jacobian at a computational cost equal to a small multiple of the cost of evaluating the function itself. To compute both columns of the Jacobian would require two forward mode evaluations.

\subsubsection{Reverse Mode}

We now perform reverse mode AD on $f$ to compute the vector-Jacobian product $\bm{\overline{y}^T J}$ at $x_1=2$ and $x_2=4$. Since $f: \mathbb{R}^2 \rightarrow \mathbb{R}$, then $\bm{J} \in \mathbb{R}^{1\times2}$. Here we will set $\bm{\overline{y}} = [1]$ which will give us the gradient \edit{$\rvect{\frac{\partial f}{\partial x_1},\frac{\partial f}{\partial x_2}}$} at $x_1=2$ and $x_2=4$:
\begin{equation}
    \overline{\bm{x}} = 
    \begin{bmatrix}
    1
    \end{bmatrix}
    \begin{bmatrix}
    \frac{\partial y}{\partial x_1} & \frac{\partial y}{\partial x_2} 
    \end{bmatrix} = \edit{
    \begin{bmatrix}
    \frac{\partial y}{\partial x_1} & \frac{\partial y}{\partial x_2}
    \end{bmatrix}}\edit{.}
\end{equation}

Reverse mode AD associates a derivative value $\overline{v}_i$ for each intermediate variable $v_i$. $\overline{v}_i$ is called the cotangent variable for $v_i$. $\overline{v}_i$ is computed after the function evaluates $y$, by combining partial derivatives from end to beginning. Because $\bm{\overline{y}} = [1]$, then the cotangent variable $\overline{v}_i \equiv \frac{\partial y}{\partial v_i}$ for each variable.

\begin{table}
    \centering
    \begin{tabular}{|c|c|c|}\hline
        Step & Variable & Value \\ \hline
        1 & $x_1$ & 2.0 \\ \hline
        2 & $x_2$ & 4.0 \\ \hline
        3 & $v_1$ & 7.389 \\ \hline
        4 & $v_2$ & 29.556 \\ \hline
        5 & $v_3$ & -0.959 \\ \hline
        6 & $v_4$ & 4.0 \\ \hline
        7 & $v_5$ & 1.0 \\ \hline
        8 & $y$ & 0.0414 \\ \hline
    \end{tabular}
    \quad
    \begin{tabular}{|c|c|c|}
    \hline
    Step & Cotangent & Value \\ \hline
    9 & $\overline{y} = \frac{\partial y}{\partial y}$ & 1.0 \\ \hline
    10 & $\overline{v}_5 = \edit{\overline{y}}\frac{\partial y}{\partial v_5}$  & 1.0 \\ \hline
    11 & $\overline{v}_4 = \overline{v}_5 \frac{\partial v_5}{\partial v_4}$ & 0.25   \\ \hline
    12 & $\overline{v}_3 = \edit{\overline{y}} \frac{\partial y}{\partial v_3}$ & 1.0 \\ \hline
    13 & $\overline{v}_2 = \overline{v}_3 \frac{\partial v_3}{\partial v_2}$ & -0.285   \\ \hline
    14 & $\overline{v}_1 = \overline{v_2}\frac{\partial v_2}{\partial v_1}$  & -1.140 \\ \hline
    15 & $\overline{x}_2 = \overline{v}_5 \frac{\partial v_5}{\partial x_2} + \overline{v}_2 \frac{\partial v_2}{\partial x_2}$ & -2.355 \\ \hline 
    16 & $\overline{x}_1 = \overline{v}_4 \frac{\partial v_4}{\partial x_1} + \overline{v}_1 \frac{\partial v_1}{\partial x_1}$ & -7.421 \\ \hline
    \end{tabular}
    \caption{The steps of the reverse mode computation which compute the function from equation \ref{eq:examplefunc} and its derivatives $\frac{\partial f}{\partial x_1}$ and $\frac{\partial f}{\partial x_2}$. In steps 1 and 2, the input variables $x_1$ and $x_2$ are set. In steps 3-8, the forward pass of the computation is performed and the value of each intermediate variable is stored for use in the backwards pass. In step 9, the cotangent variable $\overline{y}$ is set to 1. In steps 10-16, the computational graph in figure \ref{fig:compgraphAD} is traversed in reverse topological order and the cotangent variables of each variable in the graph are computed using the update rule in equation \ref{eq:reverseupdate}. }
    \label{tab:reversemode}
\end{table}

We can use the computational graph in figure \ref{fig:compgraphAD} along with table \ref{tab:reversemode} to understand the detailed computations performed by reverse mode AD. On the left side of table \ref{tab:reversemode}, the forward pass is performed. In the forward pass, the computational graph in figure \ref{fig:compgraphAD} is traversed in topological order. For each vertex, $v_i$ is computed and stored in memory. On the right side of table \ref{tab:reversemode}, the backwards pass is performed. In the backwards pass, the computational graph is traversed in reverse topological order. The cotangent variable $\overline{y}$ is first set to 1. Then, for each vertex the cotangent variable $\overline{v}_i$ is computed using the update rule 
\begin{equation}\label{eq:reverseupdate}
    \overline{v}_i \equiv \frac{\partial y}{\partial v_i} = \sum_{\substack{j \in \text{children} \\ \text{of } i}} \overline{v}_j \frac{\partial v_j}{\partial v_i}\edit{.}
\end{equation}
\edit{This update rule is equivalent to the chain rule.}

At the end of the backwards pass, we have the result of the computation $y = 0.0414$, as well as the cotangent variables $\overline{x}_1 \equiv \frac{\partial f}{\partial x_1} = - 7.421$ and $\overline{x}_2 \equiv \frac{\partial f}{\partial x_2} = - 2.355$. Notice that reverse mode AD gives the full Jacobian of a scalar function at a computational cost equal to a small multiple of the cost of evaluating the function itself.

\section{Multi-Filament Coil Parametrization}

Existing coil design codes have ultimately optimized the positions of zero-thickness or filamentary coils.
A filamentary approximation to a finite-build coil is valid in the limit that the coil thickness is much less than the minimum distance between the coils and the plasma. If the coil thickness is non-zero, which will be true of any real set of coils, then the errors in this approximation are second-order in the ratio of the coil thickness to the coil-plasma distance as shown in equation \ref{eq:expansion}. To account for these finite-build corrections, we need to directly optimize coils with non-zero thickness. 

\begin{figure}
    \centering
    \includegraphics[scale=0.1]{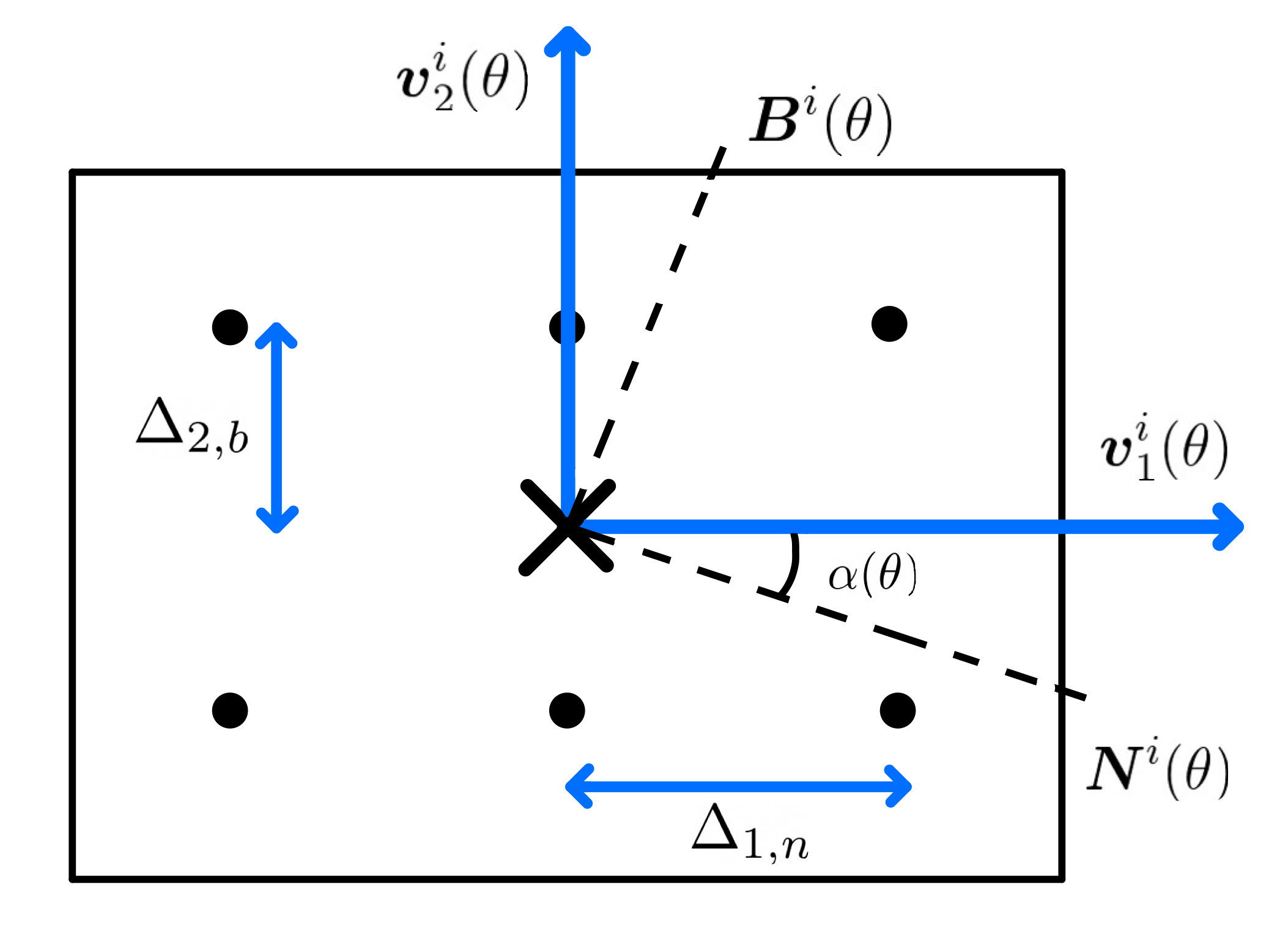}
    \caption{A coil cross-section showing the multi-filament approximation to the coil winding pack used by FOCUSADD\edit{; the coil tangent vector $\bm T^i$ (not shown) is normal to the plane}. The winding pack centroid is shown as a black X. The current-carrying filaments are shown as black circles. The filaments are placed on a rectangular grid whose coordinate axes are defined by the vectors $\bm{v}_1$ and $\bm{v}_2$; the coordinate axes are rotated by an angle $\alpha^i$ with respect to normal and binormal vectors $\bm{N}^i$ and $\bm{B}^i$. The indices of the rectangular grid are $n$ and $b$ \edit{and are counted from the bottom left corner}; filaments at index $n$ are displaced relative to the winding pack centroid in the $\bm{v}_1$ direction by a distance $\Delta_{1,n}$ and filaments at index $b$ are displaced relative to the winding pack centroid in the $\bm{v}_2$ direction by a distance $\Delta_{2,b}$.}
    \label{fig:multifilament}
\end{figure}

In this work, we directly optimize finite-build modular coils using a multi-filament approximation to the winding pack. 
\edit{The filaments used in the multi-filament approximation are, at each poloidal angle $\theta$, placed on a rectangular grid centered on the winding pack centroid $\bm r^i$.}\footnote{\edit{What we call the ``winding pack centroid", other authors have called the ``central filament". We choose this term because our multi-filament approximation may have no central current-carrying filament, see figure \ref{fig:multifilament}.}}
\edit{The position of the winding pack centroid $\bm{r}^i = \{x^i, y^i, z^i\}$ is parametrized as a function of $\theta$ with a Fourier series given by}
\begin{equation}
     \label{eq:coil_r}
     \begin{split}
        x^i(\theta) &= \sum_{m=0}^{N_F-1} \left[ X^i_{c,m} \cos(m \theta) + X^i_{s,m} \sin(m \theta) \right] \\
        y^i(\theta) &= \sum_{m=0}^{N_F-1} \left[Y^i_{c,m} \cos(m \theta) + Y^i_{s,m} \sin(m \theta) \right]\\
        z^i(\theta) &= \sum_{m=0}^{N_F-1} \left[Z^i_{c,m} \cos(m \theta) + Z^i_{s,m} \sin(m \theta) \right]
     \end{split}\edit{.}
\end{equation}
\edit{$N_F$ is an integer describing the number of modes in the Fourier series.} The parameters of this Fourier series can be combined into a single vector $\bm{R}$, where $\bm{R} \equiv \big\{\bm{X}^i_c,\bm{X}^i_s, \bm{Y}_c^i,\bm{Y}_s^i, \bm{Z}_c^i,\bm{Z}_s^i\big\}, \ i=1, \cdots, N_c$\edit{, where $N_c$ is the number of coils}.

Like the OMIC code \cite{l_singh_preprint}, we use a multi-filament approximation to the coil winding pack and have the freedom to allow the winding pack to rotate. Figure \ref{fig:multifilament} displays a cross-section of the multi-filament approximation to a coil winding pack at a particular poloidal angle $\theta$. The centroid of the coil winding pack\edit{, $\bm r^i$,} is shown with a black X in figure \ref{fig:multifilament}; the black circles\edit{, $\bm{r}^i_{n,b}$, are current-carrying filaments indexed by $n$ and $b$, starting from the bottom left corner}. In figure \ref{fig:multifilament}, six filaments are placed on a 3 by 2 grid, but any number of filaments are allowed. 

The vectors $\bm{v}^i_1(\theta)$ and $\bm{v}^i_2(\theta)$ in figure \ref{fig:multifilament} \edit{form} the coordinate axes of the rectangular grid upon which the coil filaments are placed. $\bm{v}^i_1$ and $\bm{v}_2^i$ are rotated relative to a normal vector $\bm{N}^i$ and binormal vector $\bm{B}^i$ by an angle $\alpha^i(\theta)$, given by 
\begin{equation}\label{eq:rotate_coil}
    \begin{bmatrix}
        \bm{v}^i_1(\theta) \\ 
        \bm{v}^i_2(\theta) 
    \end{bmatrix} = 
    \begin{bmatrix}
        \cos\alpha^i & -\sin\alpha^i \\
        \sin\alpha^i & \cos\alpha^i
    \end{bmatrix} 
    \begin{bmatrix}
        \bm{N}^i(\theta) \\
        \bm{B}^i(\theta) 
    \end{bmatrix}\edit{.}
\end{equation}
$\alpha^i$ is parametrized by another Fourier series, given by
\begin{equation}\label{eq:rotfourier}
    \alpha^i(\theta) = \frac{N_R \theta}{2} +  \sum_{m=0}^{N_{FR}-1
    } \left[ A_{c,m}^i \cos{(m\theta)}  + A_{s,m}^i \sin{(m\theta)} \right]\edit{.}
\end{equation}
\edit{$N_{FR}$ is an integer describing the number of modes describing this Fourier series; if the winding packs are not free to rotate it is set to zero.} $N_R$ in equation \ref{eq:rotfourier} is an integer describing the number of \edit{half} rotations of the coil winding pack; $N_R$ is normally set to zero. The parameters of this Fourier series can be combined into a single vector, $\bm{A}$ where $\bm{A} \equiv \big\{\bm{A}^i_c,\bm{A}^i_s\big\}, \ i=1, \cdots, N_c$.  \edit{The tangent vector $\bm T^i$ is defined by the Frenet-Serret equations for the winding pack centroid $\bm r^i$, while $\bm{N}^i$ and $\bm{B}^i$ are defined by the so-called ``center of mass frame", introduced in \cite{l_singh_preprint}. The center of mass frame defines the normal vector $\bm N^i$ using
\begin{equation}\label{eq:normal_vec}
\bm N^i \equiv \frac{\bm \delta^i - (\bm \delta^i \cdot \bm T^i)}{||\bm \delta^i - (\bm \delta^i \cdot \bm T^i)||},
\end{equation}
the normalized component of $\bm \delta^i$ perpendicular to $\bm T^i$, where $\bm\delta^i(\theta) \equiv \bm r^i(\theta) - \bm R^i_{c,0}$ and $\bm R^i_{c,0}$ is the center of mass of the $i$th coil. The binormal vector $\bm B^i$ is defined as $\bm T^i \times \bm N^i$. $\bm T^i(\theta)$, $\bm N^i(\theta)$, $\bm B^i(\theta)$ define an orthonormal coordinate system around each coil at each poloidal angle $\theta$; this coordinate system defines the orientation of the winding pack at zero rotation.}

The position of $\bm{r}^i_{n,b}$, the filament in the $i$th coil with indices $n$ and $b$, is given by
\begin{equation}\label{eq:spacing}
    \bm{r}^i_{n,b}(\theta) = \bm{r}^i + \Delta_{1,n} \bm{v}_1^i(\theta) + \Delta_{2,b} \bm{v}_2^i(\theta)\edit{.}
\end{equation}
$n$ and $b$ are indices of the filaments on the rectangular grid\edit{, counting up from 1}. $\Delta_{1,n} \equiv (n - \frac{N_1+1}{2})l_1$ and $\Delta_{2,b} \equiv (b - \frac{N_2+1}{2}) l_2$ where $l_1$ and $l_2$ are the spacing between gridpoints in the $\bm{v}_1$ and $\bm{v}_2$ directions and $N_1$ and $N_2$ are the number of gridpoints in the $\bm{v}_1$ and $\bm{v}_2$ directions.

In this section, we have described the mathematical model for the multi-filament coil parametrization used by FOCUSADD. Each finite-build coil is parametrized by four Fourier series, three for the centroid of the coil winding pack and one for the \edit{orientation} of the winding pack in space. The parameters of these Fourier series can be combined into a single vector $\bm{p}$, where $\bm{p} \equiv \big\{ \bm{R},\bm{A} \big\}$.

\section{Objective Function}

The goal of stellarator coil design is to find a set of coils which reproduces the target magnetic field well enough to accomplish the performance goals of the experiment and which can be built and assembled at the cheapest possible cost. In practice, this design problem is formulated as an optimization problem. The goal is to find optimal coil parameters $\bm{p}^*$ which minimize an objective function $f$. 

\begin{equation}
    \bm{p}^* = \argminC_{\bm{p}} f(\bm{p})
\end{equation}
This objective function should be chosen to meet the goals of the experiment, and therefore should incorporate both physics and engineering objectives. The standard way of incorporating multiple objectives in an optimization problem is to sum the multiple objectives into a total objective function $f_{total}$. 

\begin{equation}\label{eq:ftotal}
    f_{total}(\bm{p}) = f_{Phys}(\bm{p}) + f_{Eng}(\bm{p})
\end{equation}

Gradient-based optimization is commonly used to optimize high-dimensional non-convex objective functions. Performing gradient-based optimization requires the computation of derivatives of the objective function. Reverse mode AD allows for the gradient to be computed efficiently -- independent of the number of parameters, and at a runtime cost of \edit{a small multiple of} the cost of computing the objective function -- for arbitrary differentiable functions. In practice, AD makes gradient-based optimization of high-dimensional objective functions easy and efficient. The job of the stellarator coil designer is then to craft an objective function which accounts for the complex physics and engineering objectives required of the stellarator coils and which can be effectively optimized. 

For simplicity we choose to optimize the same objective function as FOCUS. The physics objective function is the so-called ``quadratic flux'', given by the integral over the outer toroidal surface of the square of the dot product between the magnetic field $\bm{B}$ and the surface normal vector $\bm{n}$\edit{:}
\begin{equation}\label{eq:quadratic_flux}
    f_{Phys} \equiv \int_S (\bm{B} \bm{\cdot} \bm{n})^2 dA\edit{.}
\end{equation}
For simplicity, FOCUSADD ignores the magnetic field generated due to currents in the plasma. The vacuum magnetic field is given by the Biot-Savart law integrated over the filaments in each multi-filament coil:
\begin{equation}\label{eq:biot_savart}
    \bm{B}(\bm{r}) = \sum_{i=1}^{N_c} \sum_{n=1}^{N_1} \sum_{b=1}^{N_2} \frac{\mu_0  I^i_{n,b}}{4 \pi} \oint \frac{d\bm{l}^i_{n,b} \times (\bm{r} - \bm{r}^{i}_{n,b})}{|\bm{r} - \bm{r}^{i}_{n,b}|^3}\edit{.}
\end{equation}
For simplicity, FOCUSADD assumes each filament carries the same current and does not treat each coil's current as an optimization parameter. 

Here we choose $f_{Eng}$ to be proportional to the average length of the coil centroids, as given by equation
\begin{equation}\label{eq:feng}
    f_{Eng} \equiv \lambda_L \frac{1}{N_c} \sum_{i=1}^{N_c} L_i\edit{.}
\end{equation} 
We weigh the length penalty with a regularization coefficient $\lambda_L$. Although the raw material cost of the coils will be approximately proportional to the length of the coils, this simplified engineering objective function is likely a poor approximation to the true cost of manufacturing and assembling real stellarator coils. Future coil design codes can and should account for the many complex engineering requirements on the coils; automatic differentiation makes optimizing these objective functions easy and efficient.

\begin{figure}
    \centering
    \includegraphics[scale=0.14]{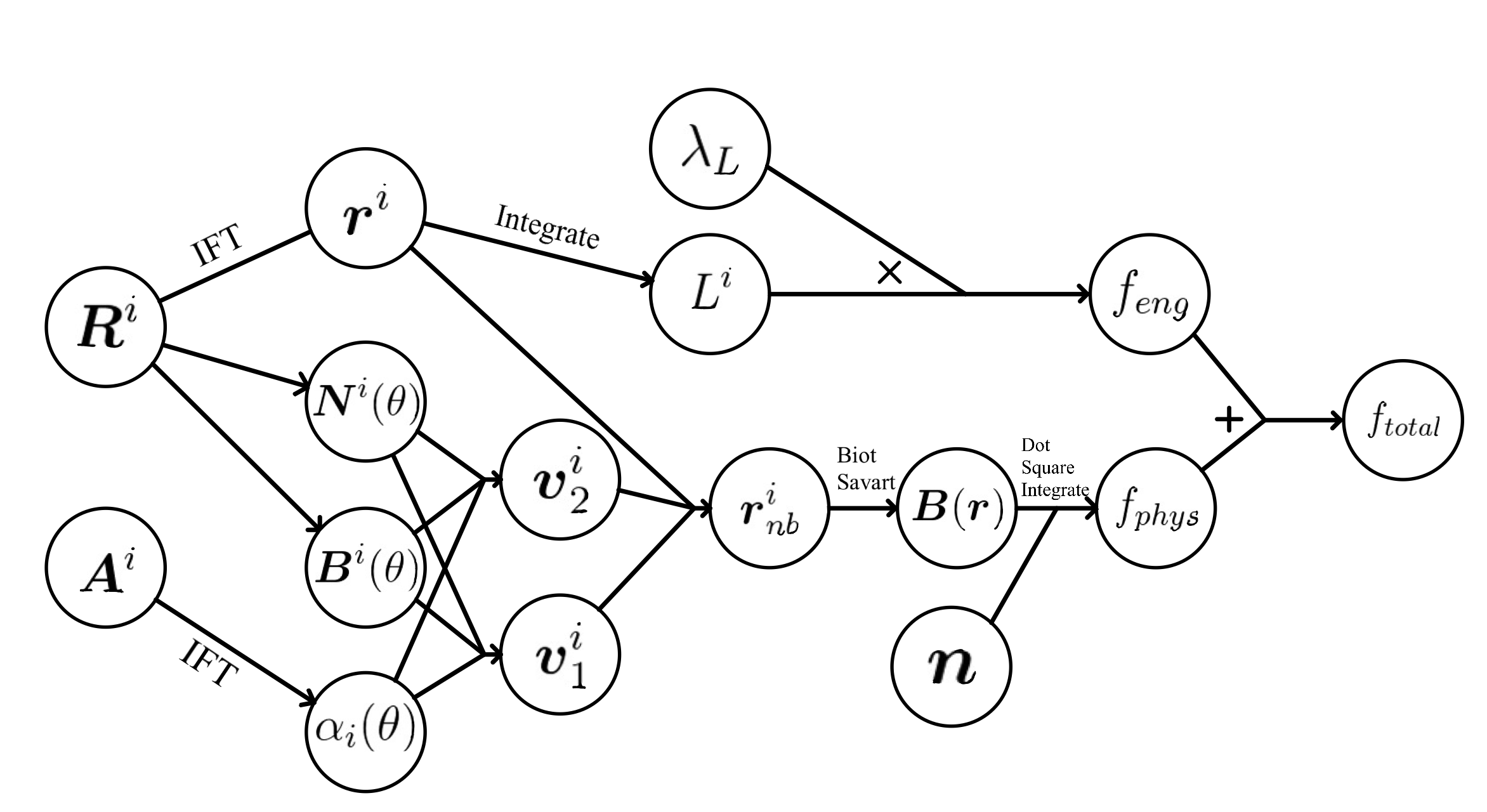}
    \caption{A computational graph for the function defined in equation \ref{eq:ftotal}. The variables on the left, $\{\bm{R}^i\}_{i=1}^{N_c}$ and $\{\bm{A}^i\}_{i=1}^{N_c}$, represent the coil parameters we want the gradient of $f_{total}$ with respect to. ``IFT" stands for ``inverse Fourier transform''. To compute this gradient, we use reverse mode AD by computing the intermediate variables in topological order and then the cotangent variables in reverse topological order. }
    \label{fig:compgraphftotal}
\end{figure}

A computational graph which represents the calculation of the objective function in equation \ref{eq:ftotal} is shown in figure \ref{fig:compgraphftotal}. The vertices are the variables computed as a result of the computation and the edges are the operations performed on those variables. Not all intermediate variables or operations are shown in figure \ref{fig:compgraphftotal}. We would like to compute the gradient of $f_{total}$ with respect to $\bm{p}\equiv \big\{ \bm{R},\bm{A} \big\}$. To compute this gradient, an AD tool computes $f_{total}$ by traversing the graph from beginning to end, then computes the cotangent variables by traversing the graph from end to beginning. At the end of the computation, we have the gradient $\frac{d f}{d \bm{p}}$. Gradient descent with momentum \edit{\cite{momentum}} is used to minimize the objective function.

\section{Finite-Build Coil Optimization Results}

To understand the effect of optimizing finite-build coils, we investigate two configurations: (i) a rotating elliptical stellarator, and (ii) a boundary for \edit{a} W7-X\edit{-like stellarator}. We first use the rotating elliptical stellarator to both establish the viability of the method and to demonstrate the \edit{optimization} of the coil winding pack \edit{orientation}. We then use the W7-X\edit{-like} boundary to test the algorithm and to investigate the effects of including finite-build on the resulting coils. \edit{In each case, the center of mass frame was used to define the zero-rotation frame, and nine filaments were placed on a square 3 by 3 grid in a multi-filament approximation to the coil finite build. Although a real conductor or superconductor will typically have more than nine filaments, adding more filaments is an increasingly small correction. For our purposes of testing the algorithm and understanding the finite-build effects, nine filaments is sufficient. }

\subsection{Elliptical Stellarator}

\begin{figure}
    \centering%
    \includegraphics[scale=0.13]{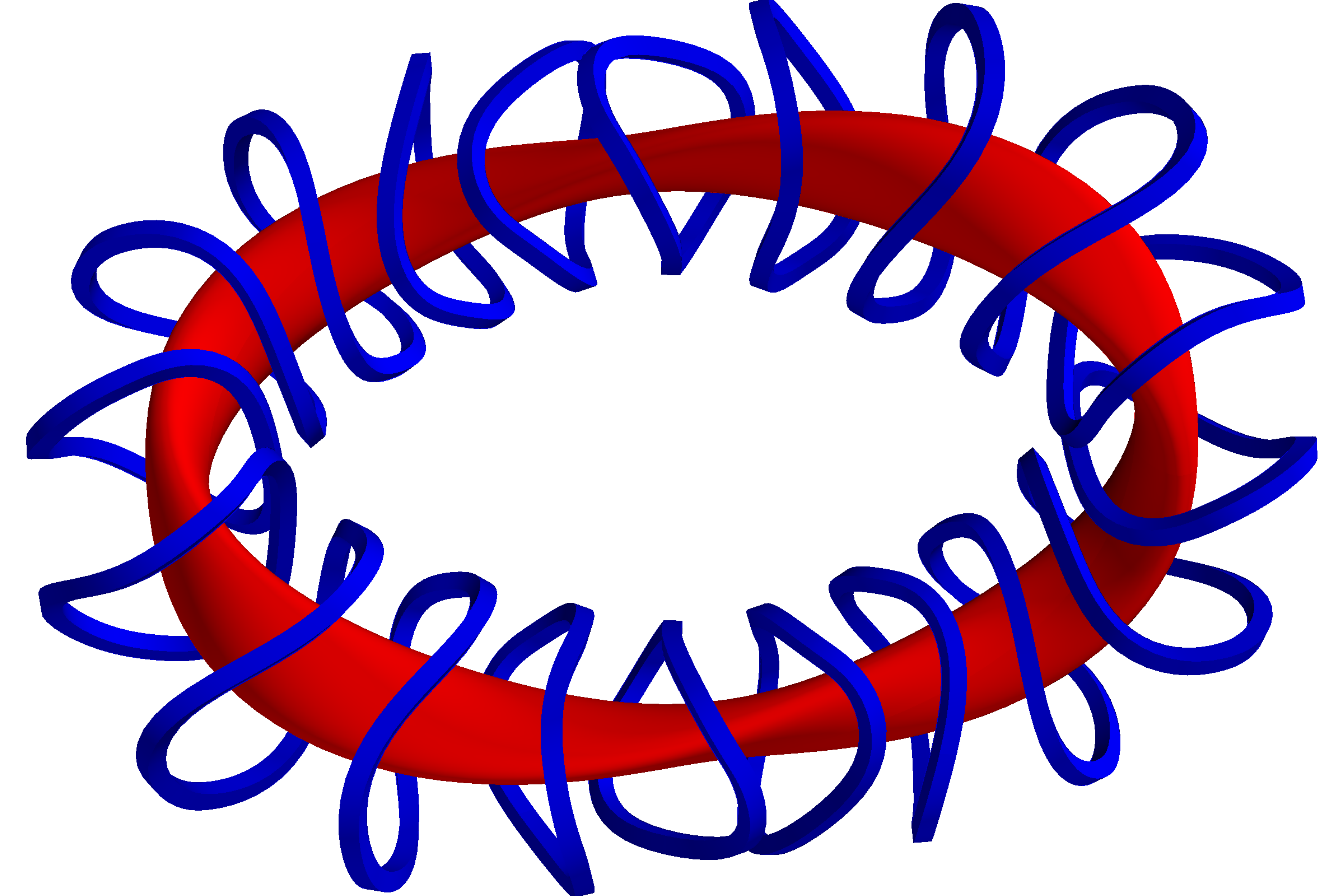}
    \caption{The outer toroidal surface and optimized zero-rotation coils of the four-period elliptical rotating stellarator used to test the FOCUSADD method. A length penalty of $\lambda_L = 0.1$ is chosen to force the coils to remain close to the plasma. The coil winding pack centroids overlay filamentary optimized coils (not shown), demonstrating that the finite-build is a small correction. }
    \label{fig:elliptical_stellarator}
\end{figure}

\begin{figure}
    \centering%
    \includegraphics[scale=0.4]{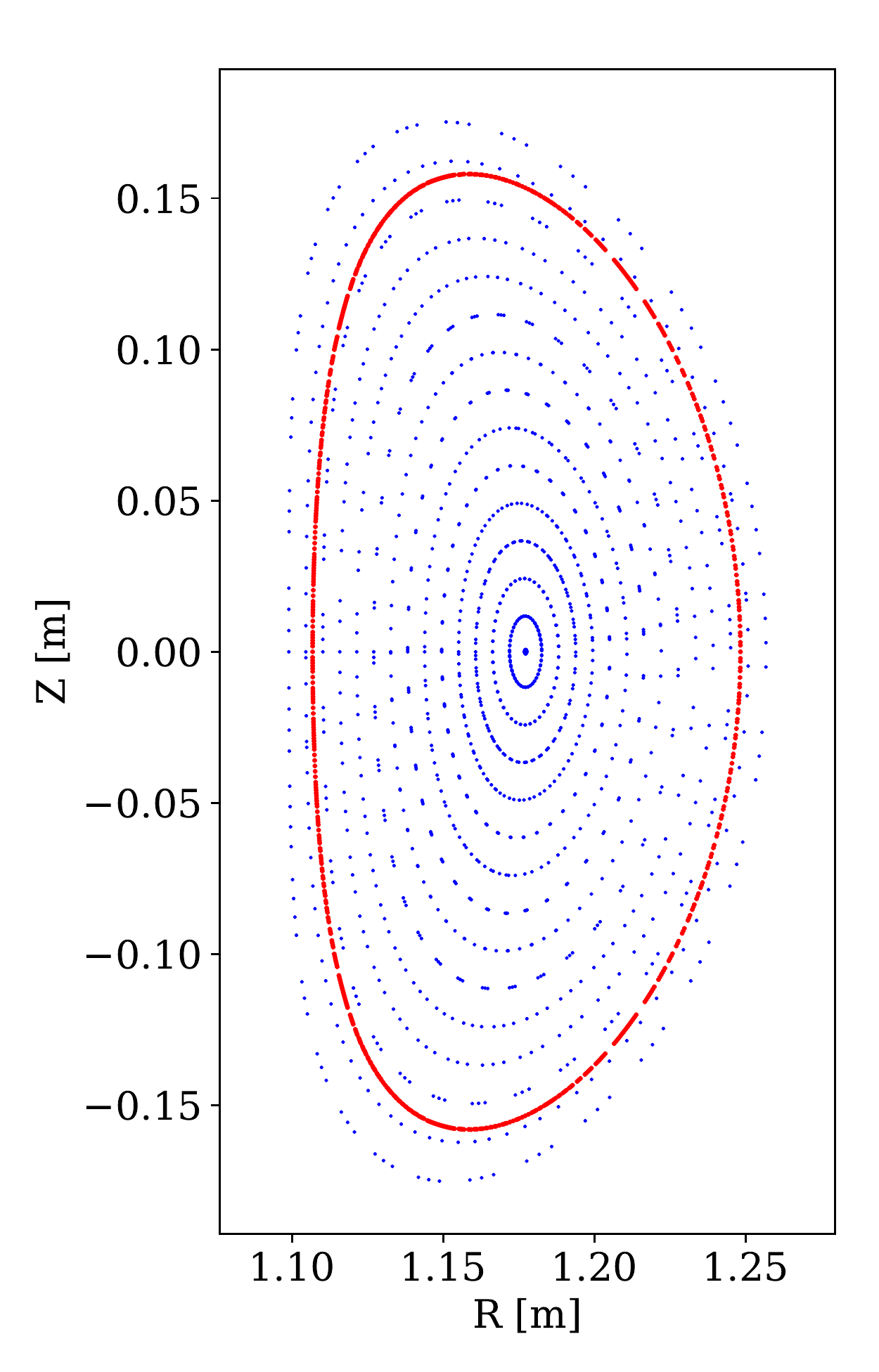}
        \caption{A poincare plot of the magnetic field lines of the optimized finite-build coils in figure \ref{fig:elliptical_stellarator}, showing nested magnetic surfaces. The target surface is shown in red. }
    \label{fig:poincare}
\end{figure}

The FOCUSADD code was used to find optimized coils for a rotating elliptical stellarator. This allowed us to establish the viability of the method and to demonstrate the \edit{optimization} of the coil winding pack \edit{orientation}. The outer surface of the four-period, elliptical cross-section stellarator is shown in figure \ref{fig:elliptical_stellarator}. Also shown in figure \ref{fig:elliptical_stellarator} are optimized finite-build coils with zero rotation of the winding pack relative to the center of mass frame. A Poincare plot showing nested magnetic surfaces for optimized finite-build coils is shown in figure \ref{fig:poincare}. 


\subsubsection{Winding Pack \edit{Orientation Optimization}}

\begin{figure}
    \centering
    \includegraphics[scale=0.19, angle=90]{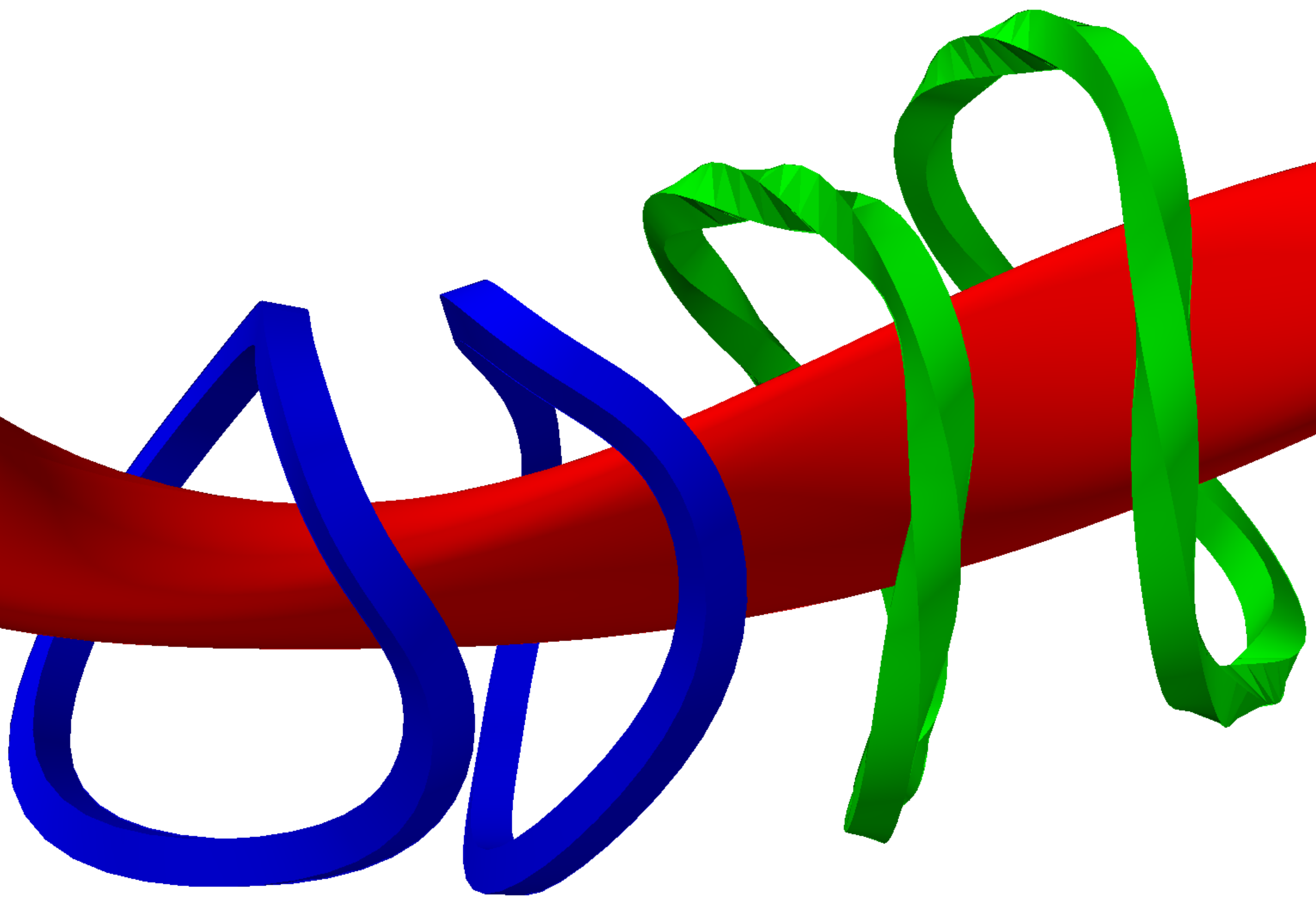}
    \caption{Finite-build coils are compared for the rotating elliptical stellarator. The leftmost two coils (blue) have zero winding pack rotation \edit{relative to the center of mass frame}, while the rightmost two coils (green) are allowed to rotate freely. There are 9 filaments which are placed on a 3 by 3 square grid.}
    \label{fig:coil_comparison}
\end{figure}

We examine the case where the coil winding pack \edit{orientation} and centroid position are simultaneously optimized. Such optimized square cross-section finite-build coils are shown in figure \ref{fig:coil_comparison}; the right two coils (green) are allowed to rotate freely while the left two coils (blue) have zero winding pack rotation. Allowing the coils to freely rotate decreases the quadratic flux by 20\% relative to the zero-rotation coils. However, due to the \edit{rapidly twisting} winding pack profile, these coils would be extremely challenging to engineer; we conclude that the winding pack cannot simply be allowed to rotate freely. This conclusion is consistent with the results from Singh, et al. \cite{l_singh_preprint}, who find that regularizing the coil rotation profiles leads to coils with less rotation relative to the center of mass frame. While a regularization penalty on the coil rotation \edit{profile} could easily be added to FOCUSADD, we have chosen not to focus on rotation profile optimization in this paper. 

The rest of this paper uses coils with zero winding pack rotation in the center of mass frame; this particular choice of frame is arbitrary but results in simple winding packs. Further discussion of the coil winding pack \edit{orientation} can be found in the conclusion.

\subsection{W7-X\edit{-Like Stellarator} }

To investigate the effect of directly optimizing finite-build coils, we find optimized coils for a W7-X\edit{-like stellarator} surface. \edit{The particular surface is geometrically similar to the standard configuration of W7-X, although no known reference for it exists in the literature. We make no attempt to exactly match the standard configuration surface.} \edit{Wendelstein 7-X (W7-X) is a 5.5m major radius, five-period experimental stellarator device in Greifswald, Germany. The W7-X magnetic field is produced by 50 non-planar and 20 planar modular superconducting coils. Each of the five periods consists of ten coils; due to stellarator symmetry there are only five unique coil shapes. We use the phrase ``W7-X-like" to emphasize that} the goal at this stage is not to find coils as if designing an actual experiment\edit{, nor is the goal to compare to any experimental design. R}ather\edit{,} the goal is to test the method and determine approximately the importance of the finite-build effects \edit{on a realistic stellarator configuration}; this goal can most simply be phrased with the question \textit{``Does finite-build matter?"}.

\edit{To investigate these finite-build effects we focus on two quantities. The first quantity, $\Delta r$, is (roughly speaking) the distance the coils shift due to finite-build. $\Delta r$ is defined as the mean minimum distance between the winding pack centroid of optimized zero-rotation finite-build coils and optimized filamentary coils. The second quantity, $\Delta e$, is (roughly speaking) by what percent does the normalized field error change if finite-build is not accounted for in the optimization. The normalized field error $e$ is defined as
\begin{equation}\label{eq:normalized_field_error}
    e \equiv \frac{1}{\int_S dA }\int_S \frac{|\bm B \cdot \bm n|}{|B|} dA.
\end{equation}
$\Delta e \equiv (\nicefrac{e'_{fil}}{e_{fb}} - 1) * 100$ is defined as the ratio of $e'_{fil}$, the normalized field error of coils which are optimized with a filamentary approximation but then are built with zero rotation and finite-build, and $e_{fb}$, the normalized field error of zero-rotation coils which are optimized with a finite-build approximation and then built with zero rotation and finite-build, minus 1, all times 100.}
\begin{figure}
    \centering
    \includegraphics[scale=0.16]{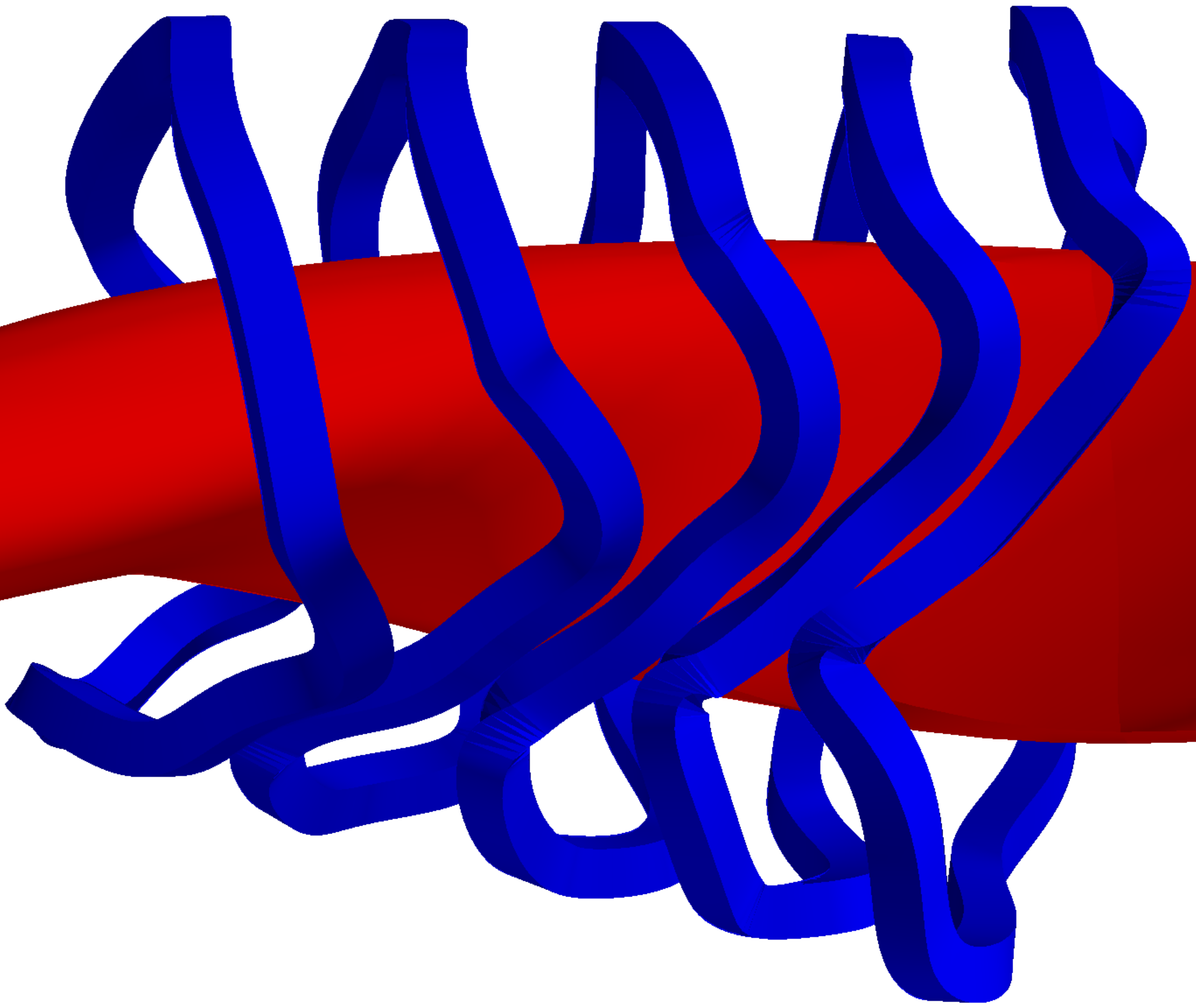}
    \caption{Half-period consisting of five W7-X finite-build coils. Coils shown are $18$cm by 18cm in cross-section.}
    \label{fig:w7x}
    \vspace{-0.5cm}%
\end{figure}

A half-period of a W7-X\edit{-like} outer plasma surface and optimized zero-rotation finite-build coils are shown in figure \ref{fig:w7x}. The finite-build coil dimensions (18cm by 18cm) in figure \ref{fig:w7x} are approximately the same dimensions as the true W7-X conductor (\edit{19.2cm by 16cm}) \edit{\cite{w7x_coilsize}}. The regularization penalty on the coil lengths, $\lambda_L$, is set to 0.25. \edit{Each optimization run is performed for 10,000 iterations, which ensures convergence.}

\begin{figure}
        \centering%
        \begin{subfigure}{0.49\textwidth}
        \centering%
        \includegraphics[scale=0.4]{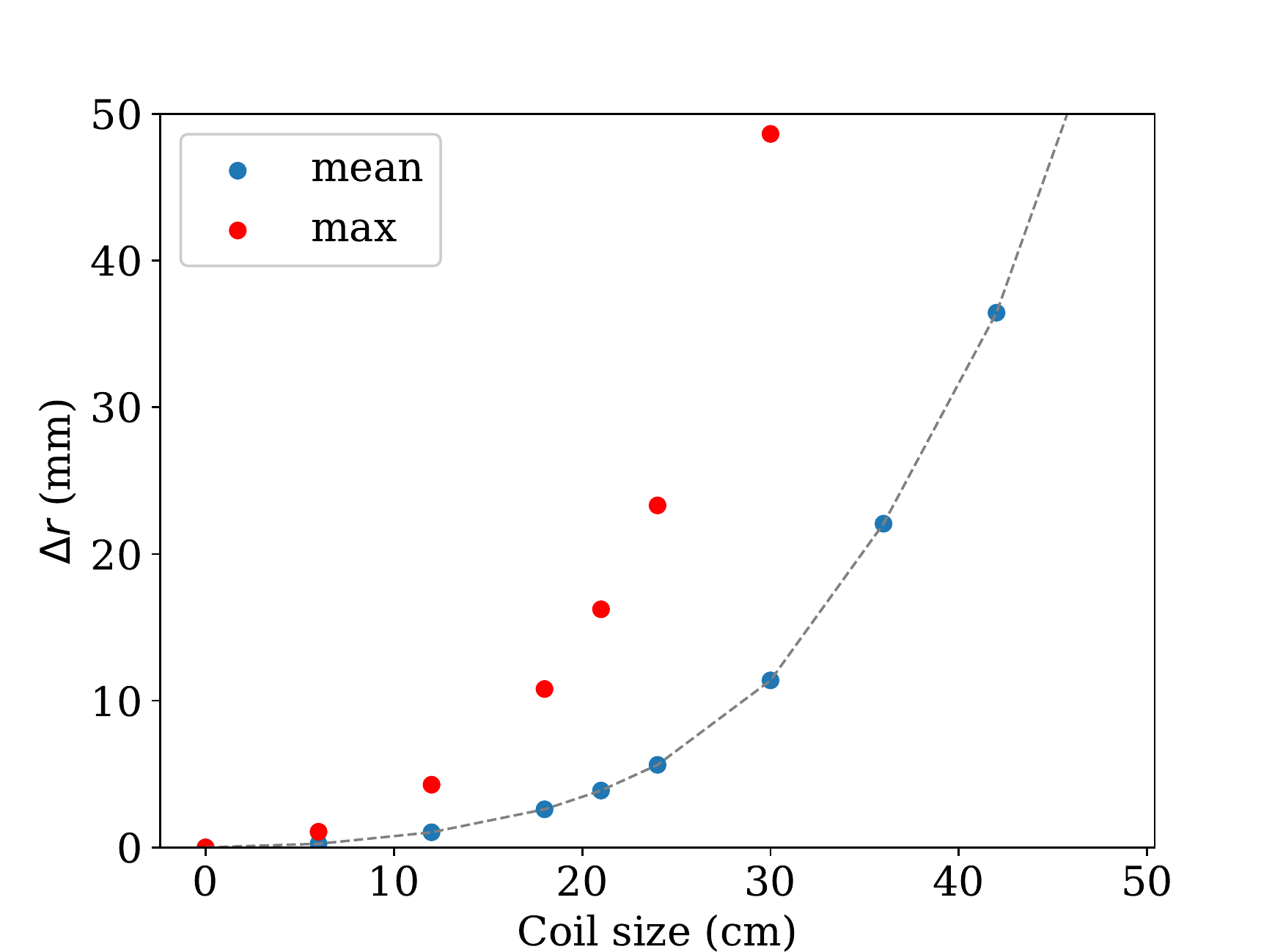}
        \caption{$\Delta r(\delta)$ for W7-X.}
        \label{fig:w7x_r}
        \end{subfigure}
        \begin{subfigure}{0.49\textwidth}   
        \centering%
        \includegraphics[scale=0.4]{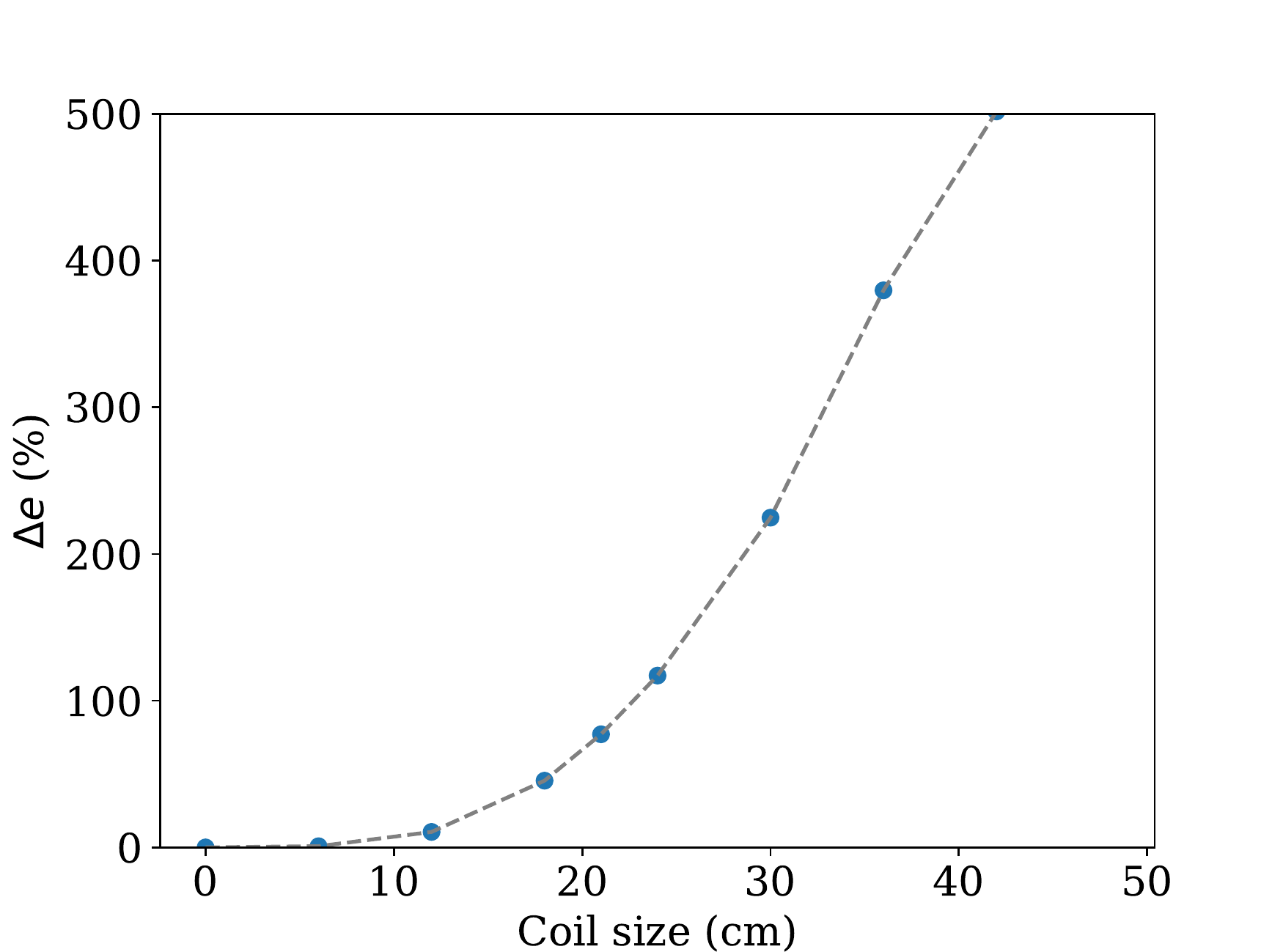}
        \caption{$\Delta e(\delta)$ for W7-X.}
        \label{fig:w7x_e}
        \end{subfigure}
        \caption{W7-X: (a) $\Delta r$, the mean shift in the coil position between finite-build and filamentary coils, is plotted in blue as a function of coil size $\delta$. The maximum shift is plotted in red. (b) $\Delta e$, the percent change in the normalized field error if finite-build is not accounted for, as a function of coil size $\delta$. For 18cm by 18cm coils, of comparable size to the W7-X conductor cross-section (\edit{19.2cm by 16cm}), the mean shift $\Delta r \approx 2.5$mm, the maximum shift is about 11mm, and the normalized field error $\Delta e \approx 50\%$.}
        \label{fig:curves_w7x}
        \vspace{-0.5cm}%
    \end{figure}

To investigate the effect\edit{s} of including finite build, we vary the size of the $\delta$ by $\delta$ \edit{winding pack} cross-section from $\delta = 0$\edit{cm} to $\delta = 50$cm while keeping all \edit{other} quantities fixed and calculate the effects on $\Delta r$ and $\Delta e$\edit{; the results} are shown in figures \ref{fig:w7x_r} and \ref{fig:w7x_e}. \edit{Although eventually the coils will overlap as $\delta$ increases, this is not an issue for our purpose of better understanding the finite build effects. We extend $\delta$ to such a large value for the purposes of illustration.}

In figure \ref{fig:w7x_r}, we can see that as the coil size $\delta$ is increased, the mean shift in the coil positions increases. For coils of approximately the same dimensions as the W7-X coils (18cm by 18cm), we have the result that the mean shift $\Delta r \approx 2.5$mm, while the maximum shift is about 11mm. The coil tolerances in the W7-X experimental design \edit{were} \edit{$\pm 3$mm during the manufacturing process and $\pm 2$mm during the assembly process}\edit{; once the machine was assembled, the maximum deviation of any reference point was measured to be 5.7mm from its manufacture value} \cite{Andreeva_2015}. Because the mean coil shift is of the same magnitude as the coil tolerances, and the maximum shift much larger, we can reasonably conclude that finite-build effects should be accounted for in the optimization of coils for a stellarator such as W7-X.

\begin{figure}
        \centering
        \begin{subfigure}{0.45\textwidth}
        \centering
        \includegraphics[scale=0.38]{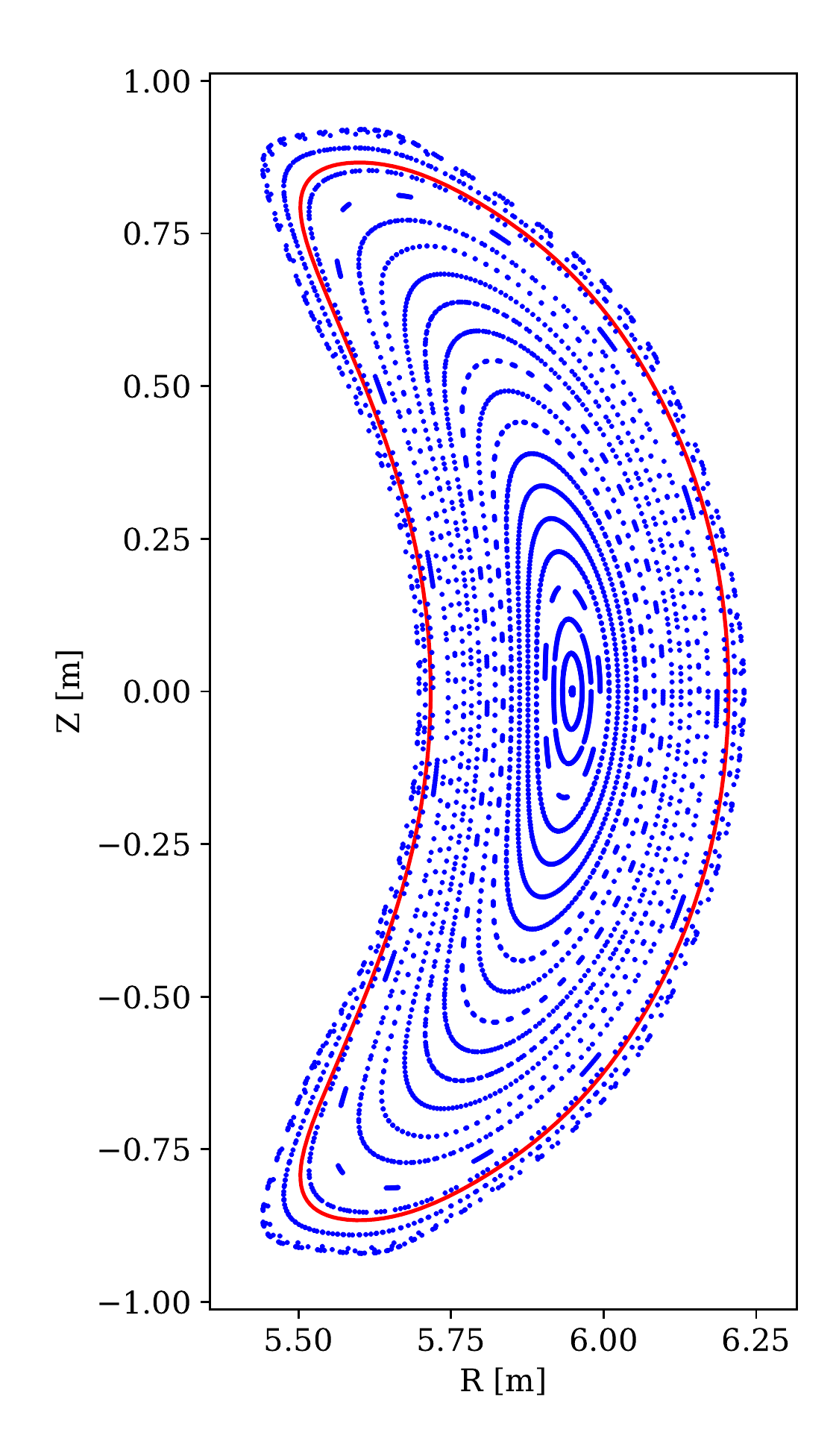}
        \caption{}
        \label{fig:poincare_fb}
        \end{subfigure}
        \begin{subfigure}{0.45\textwidth}
        \centering
        \includegraphics[scale=0.38]{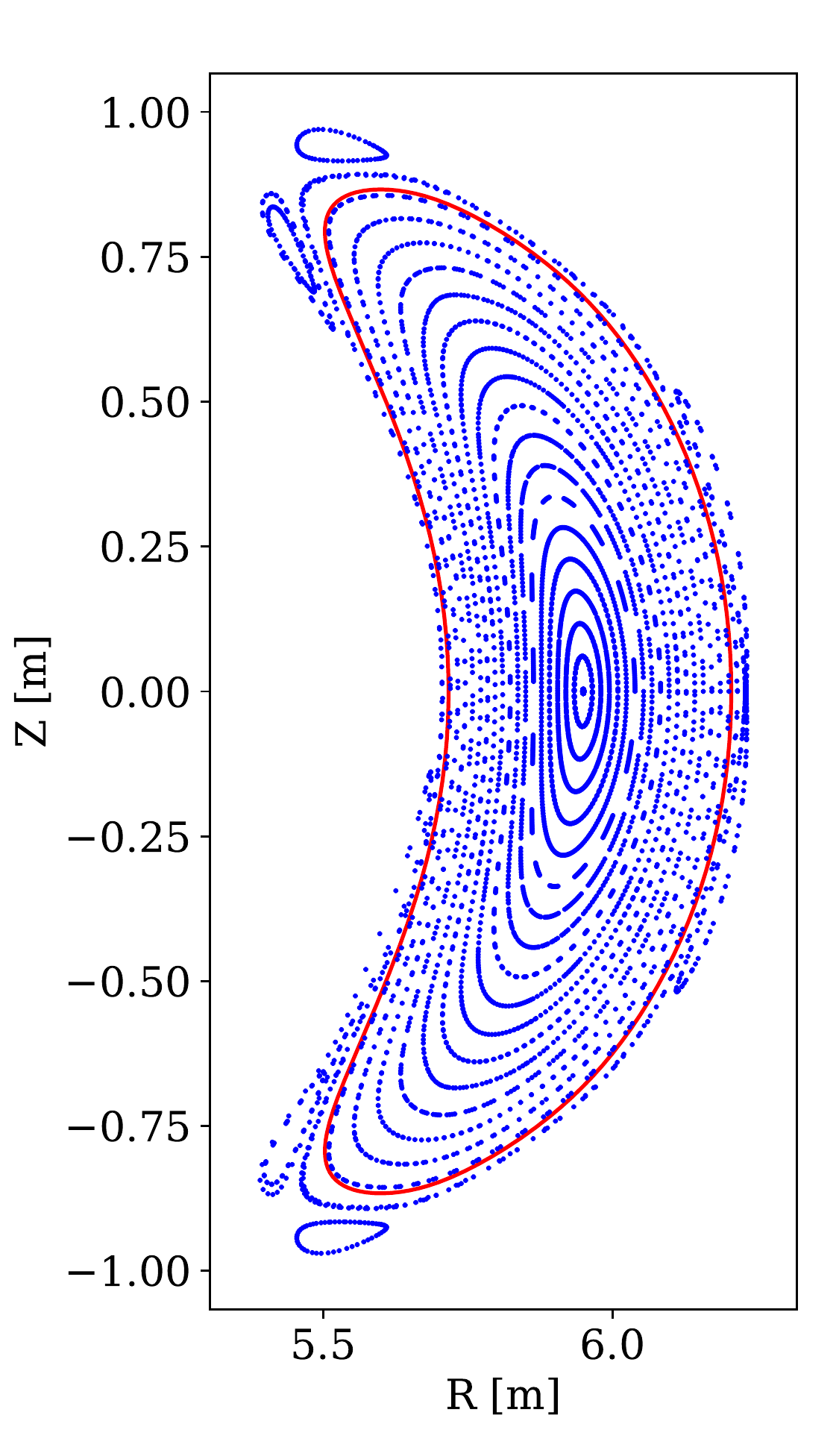}
        \caption{}
        \label{fig:poincare_18}
        \end{subfigure}
         \vskip\baselineskip
        \begin{subfigure}{0.45\textwidth}
        \centering
        \includegraphics[scale=0.38]{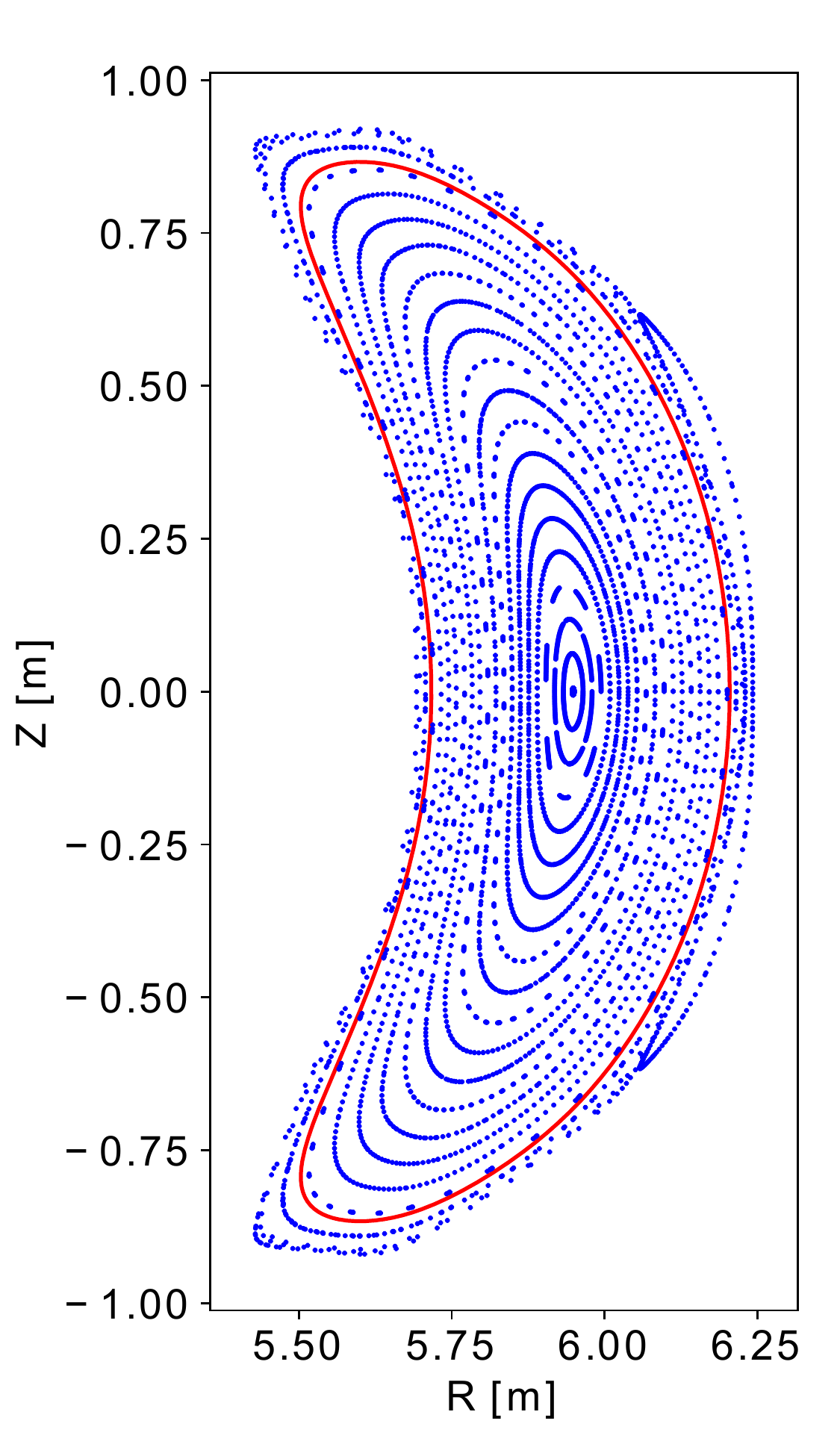}
        \caption{}
        \label{fig:poincare_fb36}
        \end{subfigure}
        \begin{subfigure}{0.45\textwidth}
        \centering
        \includegraphics[scale=0.38]{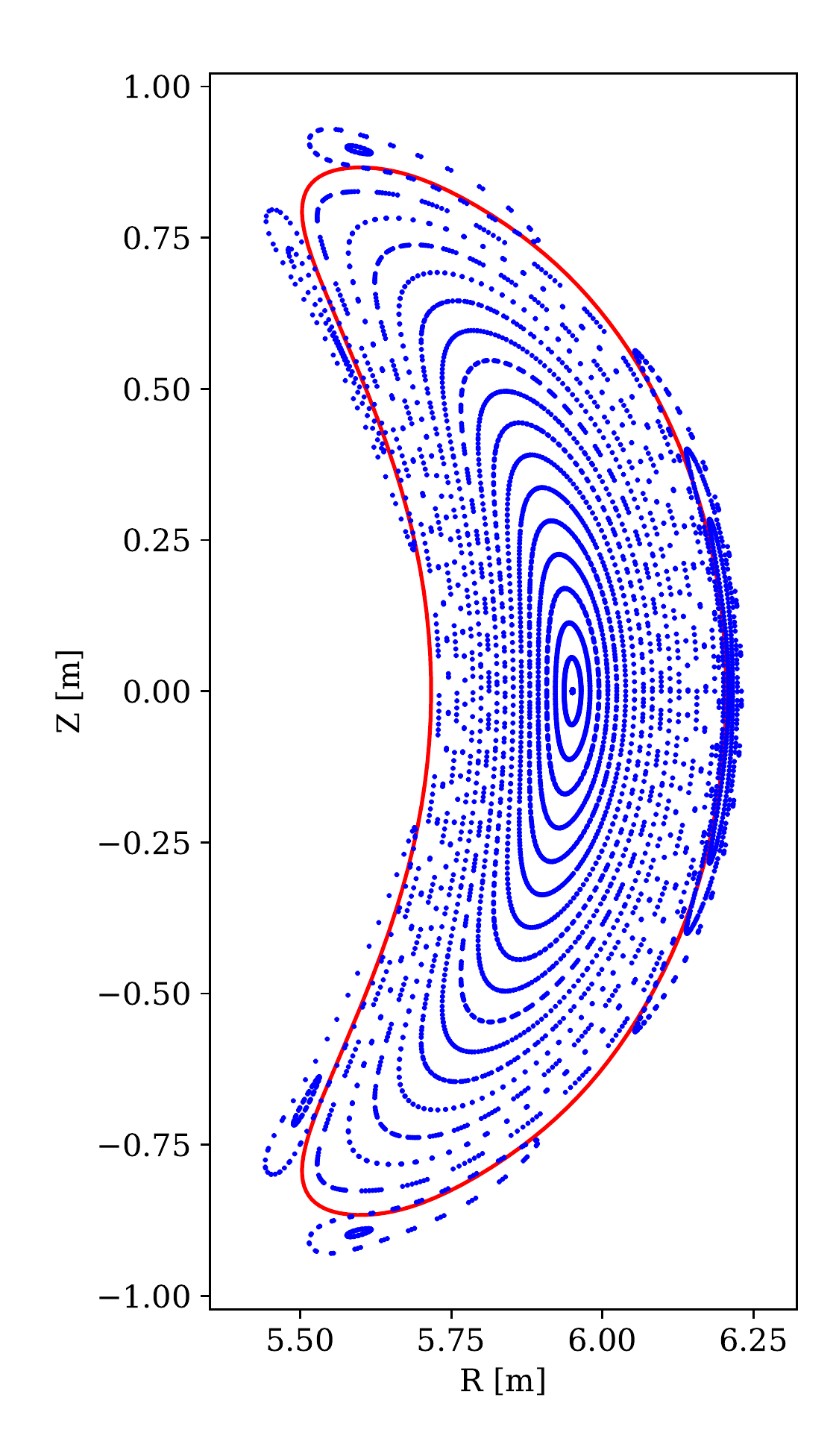}
        \caption{}
        \label{fig:poincare_36}
        \end{subfigure}
        
        \caption{Poincare plots for the W7-X\edit{-like} stellarator surface, with (a) finite build coils (18cm by 18cm), optimized using a multi-filament approximation. (b) finite build coils (18cm by 18cm), optimized using a filamentary approximation. \edit{(c) finite build coils (36cm by 36cm), optimized using a multi-filament approximation.} (d) finite build coils (36cm by 36cm), optimized using a filamentary approximation. Comparing (b) to (a) \edit{as well as (d) to (c)}, we see that optimizing finite-build coils with a filamentary approximation leads to a different island structure relative to directly optimizing finite-build coils; for this particular surface, as the field error increases, the (5,5) magnetic islands move radially inwards towards the target surface. }
        \label{fig:poincare_w7x}
\end{figure}

In figure \ref{fig:w7x_e}, we similarly see that as the coil size $\delta$ is increased, that $\Delta e$ increases. For coils of approximately the same dimensions as the W7-X coils, the normalized field error $e$ increases by approximately 50\% \edit{from its initial value of $4 * 10^{-4}$} if the finite-build is not accounted for. Determining whether this 50\% increase will lead to a significant degradation in the properties we care about (such as radial transport, MHD stability, energetic particle confinement, etc) would require a detailed analysis of the equilibrium that is outside the scope of this paper. 

\edit{Lastly, we look at} Poincare plots of the magnetic field lines \edit{to get a qualitative} understand\edit{ing} \edit{of how finite-build effects might affect the magnetic field}. 
Figure\edit{s} \ref{fig:poincare_fb} \edit{and \ref{fig:poincare_fb36}} show Poincare plots of zero-rotation finite-build coils optimized using the multi-filament model; \edit{figure \ref{fig:poincare_fb} has an 18cm by 18cm cross-section, while figure \ref{fig:poincare_fb36} has a 36cm by 36cm cross-section}. Figures \ref{fig:poincare_18} and \ref{fig:poincare_36} show Poincare plots of the magnetic field produced by finite-build coils which are optimized using a filamentary approximation, figure \ref{fig:poincare_18} has an 18cm by 18cm cross-section and $\Delta e \approx 50\%$, while figure \ref{fig:poincare_36} has a 36cm by 36cm cross-section and $\Delta e \approx 400\%$. Comparing figure \ref{fig:poincare_18} to \ref{fig:poincare_fb} \edit{as well as figure \ref{fig:poincare_36} to \ref{fig:poincare_fb36}}, we see a small qualitative difference in the magnetic field structure from finite-build coils which are optimized using a filamentary approximation. 
\edit{The structure of the magnetic field in the core shows little change, which is consistent with the understanding that finite-build effects diminish in magnitude with distance from the coils. The structure of the magnetic field in the edge shows some small changes in the size of the (5,5) magnetic island and the chaotic zones. This is also consistent with our understanding: closer to the coils, the finite build effects are larger. 
Also, because the width of the magnetic islands is proportional to the square root of the magnitude of the resonant perturbation and inversely proportional to the shear \cite{ISI:A1991FD80100018}, and overlapping islands creates chaotic fieldlines \cite{ISI:A1979HD59000001}, we expect that small changes in the magnetic field can result in noticeable changes in the island size and structure of the magnetic field in the edge region.
This} provides further evidence for the conclusion that finite build effects should be accounted for in the optimization \edit{of coils for} a stellarator such as W7-X. \edit{Finite-build effects are an important consideration for the optimization of the divertor, particularly for low-shear stellarators.}

\section{Conclusion}

We have introduced a new finite-build stellarator coil design code called FOCUSADD which represents coils as filamentary closed curves in space surrounded by a multi-filament approximation to the coil winding pack. The optimization parameters are the Fourier series describing the winding pack centroid positions and optionally the winding pack \edit{orientation}. These parameters are optimized using a simple gradient descent with momentum algorithm. The required derivatives are calculated using the reverse mode automatic differentiation (AD) tool JAX. By using reverse mode AD, the derivatives are calculated easily and efficiently, at a cost independent of the number of parameters. 

We found that by allowing coil winding packs to rotate freely in space, the quadratic flux can be further decreased, but at the cost of finding coils whose rotation profiles would be extremely difficult to engineer. Singh et al. \cite{l_singh_preprint} come to the same conclusion, but find that regularizing the coil rotation \edit{profiles} results in more feasible coils. However, each of these approaches uses the arbitrary center of mass frame; neither is based on a rigorous or quantitative understanding of what makes a coil challenging or expensive to build. Future work in finite-build stellarator coil design should develop such understanding to design either a new winding pack frame or a penalty on the rotation profile which would result in a winding pack rotation profile which is easier to engineer. An open question in finite-build stellarator coil design is whether there will be any freedom in the rotation profile to optimize for physics objectives, or whether engineering feasibility alone will determine the \edit{winding pack orientation}. 

We compared zero-rotation finite-build coils to filamentary coils for a W7-X\edit{-like} surface, and found that the optimized finite-build coils were shifted relative to the optimized filamentary coils. For coils of approximately the same sizes as were built in the experiment, the mean shift in coil positions was approximately 2.5mm. We also found that optimizing filamentary rather than finite-build coils leads to a\edit{n increase in the normalized field error $e$ as well as a} qualitatively different magnetic island structure. These calculations suggest that the finite build of stellarator coils is a non-negligible effect which should be included in the optimization of coils for real stellarator experiments.

Automatic differentiation has allowed us to compute the gradient of our objective function efficiently and easily. Although this objective function consists of the standard quadratic flux plus a simple penalty term, this objective function could easily be modified in future work to include other objective functions, such as coil-coil spacing, inter-coil electromagnetic forces, curvature, or torsion. 

\setcounter{secnumdepth}{0} 
\section{Acknowledgements}

The authors would like to thank Judith Swan, Matt Landreman, and Elizabeth Paul for helpful comments on an earlier version of this manuscript. The work is supported via U.S. DOE contract DE-AC02-09CH11466 for the Princeton Plasma Physics Laboratory.


\bibliographystyle{unsrt}

\end{document}